\documentclass[a4paper,fleqn]{cas-dc}

\usepackage{cite}
\usepackage{mathtools}
\usepackage{amsmath,amssymb,amsfonts,amsthm}
\usepackage{mathrsfs}  
\usepackage{upgreek} 
\newtheorem{theorem}{Theorem}[section]

\newtheorem{remark}{Remark}[section]

\newtheorem{assumption}{Assumption}
\usepackage{graphicx}
\usepackage[normalem]{ulem}
\usepackage{multirow}
\usepackage{booktabs}
\usepackage{enumitem}  
\usepackage{algorithm}
\usepackage{color}
\usepackage{algpseudocode}
\usepackage{threeparttable}

\usepackage{hyperref}
\usepackage{cleveref}
\usepackage{cancel}
\usepackage{subfigure}
\crefname{equation}{}{}
\Crefname{equation}{}{}
\Crefname{figure}{Fig.}{Figs.}
\Crefname{assumption}{Assumption}{Assumptions}
\crefname{assumption}{assumption}{assumptions}
\hypersetup{hidelinks=true}
\usepackage{textcomp}
\usepackage[numbers, sort&compress]{natbib}  
\def\tsc#1{\csdef{#1}{\textsc{\lowercase{#1}}\xspace}}
\tsc{WGM}
\tsc{QE}
\tsc{EP}
\tsc{PMS}
\tsc{BEC}
\tsc{DE}
\begin{document}
\let\WriteBookmarks\relax
\def\floatpagepagefraction{1}
\def\textpagefraction{.001}
\shorttitle{Relaxing Probabilistic Latent Variable Models' Specification via Infinite-Horizon Optimal Control}
\shortauthors{Zhichao Chen et~al.}

\title [mode = title]{Relaxing Probabilistic Latent Variable Models' Specification via Infinite-Horizon Optimal Control}                      
% \tnotemark[1,2]

% \tnotetext[1]{ This work was supported in part by the National Natural Science Foundation of China (NSFC) under Grant 62473103. This work was also supported in part by the National Science and Technology Major Project of China under Grant No. 2022ZD0120001, Jiangsu Provincial Scientific Research Center of Applied Mathematics under Grant No. BK20233002 %(Corresponding author: Zhiqiang Ge, Zhihuan Song)
%     .}
% \tnotetext[2]{Corresponding Authors.}

% \tnotetext[2]{The second title footnote which is a longer text matter
%    to fill through the whole text width and overflow into
%    another line in the footnotes area of the first page.}
\author{Zhichao Chen}
%\credit{Conceptualization, Methodology, Software, Validation, Writing – Original Draft}
\author{Hao Wang}
%\credit{Formal analysis, Writing – Original Draft}
\author{Licheng Pan}
%\credit{Formal analysis, Writing – Original Draft}
\author{Yiran Ma}
% %\credit{Formal analysis, Writing – Original Draft}
\author{Yunfei Teng}%\credit{Formal analysis}%[orcid=0000-0001-5785-0741]
\author{Jiaze Ma}
%\credit{Formal analysis, Investigation, Writing – Original Draft}
\author{Le Yao}
%\credit{Investigation}
\author{Zhiqiang Ge}
%\credit{Funding acquisition, Supervision, Writing – Review \& Editing}\cormark[2]
% https://orcid.org/0000-0003-4098-6479
\author{Zhihuan Song}%\credit{Funding acquisition, Supervision, Writing – Review \& Editing}\cormark[2]%\tnotemark[1]\cormark[2]

\nonumnote{Zhichao Chen, Hao Wang, Licheng Pan, Yiran Ma, and Zhihuan Song are with the State Key Laboratory of Industrial Control Technology, College of Control Science and Engineering, Zhejiang University, Hangzhou 310027, China.
  }
  \nonumnote{Yunfei Teng is with the School of Intelligence Science and Technology, Peking University, Beijing 100871, China.}
  \nonumnote{
    % Jiaze Ma is with the Department of Systems Engineering, City University of Hong Kong, Hong Kong SAR, China (e-mail: jiazema@cityu.edu.hk). 
    Jiaze Ma is with the Department of Systems Engineering, City University of Hong Kong, Hong Kong SAR, China.
    }
    \nonumnote{Le Yao is with the School of Mathematics, Hangzhou Normal University, Hangzhou 311121, China.
    }
    \nonumnote{Zhiqiang Ge is with School of Mathematics, Southeast University, Nanjing 210096, China. }
    \nonumnote{Zhihuan Song is also with the Guangdong Provincial Key Laboratory of Petrochemical Equipment Fault Diagnosis, Guangdong University of Petrochemical Technology, Maoming 525000, China.}
\begin{abstract}
In this paper, we address the issue of model specification in probabilistic latent variable models (PLVMs) using an infinite-horizon optimal control approach. Traditional PLVMs rely on joint distributions to model complex data, but introducing latent variables results in an ill-posed parameter learning problem. To address this issue, regularization terms are typically introduced, leading to the development of the expectation-maximization (EM) algorithm, where the latent variable distribution is restricted to a predefined normalized distribution family to facilitate the expectation step. To overcome this limitation, we propose representing the latent variable distribution as a finite set of instances perturbed via an ordinary differential equation with a control policy. This approach ensures that the instances asymptotically converge to the true latent variable distribution as time approaches infinity. By doing so, we reformulate the distribution inference problem as an optimal control policy determination problem, relaxing the model specification to an infinite-horizon path space. Building on this formulation, we derive the corresponding optimal control policy using the Pontryagin's maximum principle and provide a closed-form expression for its implementation using the reproducing kernel Hilbert space. After that, we develop a novel, convergence-guaranteed EM algorithm for PLVMs based on this infinite-horizon-optimal-control-based inference strategy. Finally, extensive experiments are conducted to validate the effectiveness and superiority of the proposed approach.
\end{abstract}

% \begin{graphicalabstract}
% \includegraphics{figs/cas-grabs.pdf}
% \end{graphicalabstract}

% \begin{highlights}
% \item Research highlights item 1
% \item Research highlights item 2
% \item Research highlights item 3
% \end{highlights}

\begin{keywords}
Probabilistic Latent Variable Model \sep Expectation Maximization Algorithm \sep Infinite-Horizon Optimal Control \sep Reproducing Kernel Hilbert Space.
% quadrupole exciton \sep polariton \sep \WGM \sep \BEC
\end{keywords}

\maketitle

\section{Introduction}

Probabilistic Latent Variable Models (PLVMs)~\cite{murphy2022probabilistic,cheng2025empowering,bishop1998latent}, along with their various extensions, have garnered significant interest in the automatic control community~\cite{kong2022latent,9080620}, particularly in applications such as process monitoring~\cite{yu2022generalized, raveendran2018process} and inferential sensor development~\cite{PhyscialPSFA}. These models are valued for capturing complex data distributions by representing them as more tractable joint distributions over an expanded variable space~\cite{bishop2006pattern}. Despite PVLMs' promise, improving the performance of PLVMs in downstream application scenarios remains challenging, primarily due to the inherent model specification of the latent variable distribution during the training phase.

The inclusion of the latent variable in PLVMs increases the degrees of freedom in the model's learning objective, leading to an ill-posed problem~\cite{hofmann2013regularization}. Specifically, this ill-posedness arises because there are infinitely many potential distributions for the latent variable, which complicates the optimization process. To address this, regularization terms are introduced into the learning objective to help constrain the solution space. This introduces the need for an effective parameter learning strategy. A widely used parameter estimation method for PLVMs is the Expectation-Maximization (EM) algorithm~\cite{dempster1977maximum}, which is structured as a two-step procedure: 1) the expectation step, where the latent variable distribution is inferred via an approximation distribution based on fixed model parameters and the regularization term, and 2) the maximization step, where the inferred latent variables are fixed, and the model parameters are updated. However, to reduce the computational complexity of the expectation step, the approximation distribution is typically chosen from a predefined normalized family. While this \emph{model specification} helps with computational efficiency~\cite{8588399}, it comes at the cost of reduced approximation accuracy, which hinders the overall performance of PLVMs on downstream tasks.

In this paper, we consider relaxing the model specification by leveraging the concept of infinite-horizon optimal control. Specifically, we first represent the variational distribution, a scalar function, using a finite set of particles that can be placed anywhere in the real number domain. However, the initial placement of these particles may not be optimal. To address this, we introduce an ordinary differential equation (ODE) to govern the evolution of these particles over time. Crucially, we establish that this ODE-driven particle system provides a weak solution to the continuity equation, the fundamental partial differential equation (PDE) that describes distributional flows. This theoretical connection allows us to design the particle dynamics by controlling the velocity field of the PDE. Based on this insight, we frame the inference problem as an infinite-horizon optimal control problem, where the control policy steers the particles as close as possible to the true latent variable distribution. We solve this control problem using Pontryagin's maximum principle, thereby transforming the latent variable inference problem—traditionally addressed by approximating the distribution within a predefined normalized family—into an infinite-horizon optimal control problem, which naturally relaxes model specification by inferring the optimal control policy within an infinite-horizon path space.
% In this paper, we consider relaxing the model specification by leveraging the concept of infinite-horizon optimal control. Specifically, we first represent the variational distribution, a scalar function, using a finite set of instances that can be placed anywhere in the real number domain. However, the initial placement of these instances may not be optimal. To address this, we introduce an ordinary differential equation (ODE) to perturb the instances and frame the problem as an infinite-horizon optimal control problem. This formulation aims to drive the instances as close as possible to the true latent variable distribution, which we solve using Pontryagin's maximum principle. By doing so, we transform the latent variable inference problem—traditionally addressed by approximating the distribution within a predefined normalized family—into an optimal control problem. This transformation naturally relaxes model specification by inferring the optimal control policy within an infinite-horizon path space.

Additionally, we observe that implementing optimal control problem by computer language requires real-time estimation of the intractable probability density function of the variational distribution. To overcome this challenge, we derive a closed-form solution for implementing the optimal control problem using reproducing kernel Hilbert space (RKHS). Finally, we propose a new EM algorithm based on this infinite-horizon optimal control approach for latent variable inference, prove the convergence of the proposed EM algorithm, and validate its effectiveness through comprehensive experimental evaluations.

\noindent\textbf{Contributions:} Based on the preceding contents, we summarize the key contributions of this paper as follows:
\begin{enumerate}[leftmargin=*] 
    \item{We introduce an infinite-horizon optimal control approach for inferring the latent variable distribution in PLVMs, effectively addressing the model specification issue that arises during the learning process.} 
\item{We derive a closed-form solution for implementing the optimal control problem by computer language, which bypasses the explicit estimation of the intractable variational distribution. Additionally, we propose a novel EM algorithm with proven convergence guarantees for PLVM training.}
   %  \item{We derive the closed-form expression to implement the optimal control problem, which sidesteps the explicitly estimation of the intractable variational distribution, and propose a novel convergence-guaranteed EM algorithm for PLVM training.  }
   \item{We conduct comprehensive experiments to validate the effectiveness of the proposed infinite-horizon optimal control-based latent variable distribution inference strategy and the EM algorithm.}
\end{enumerate}

\noindent\textbf{Organization:} The rest of this article is organized as follows: To demonstrate the technical gap, we first conduct a literature review in~\Cref{sec:RelatedWorks}, and to better understand this paper, we introduce preliminaries in~\Cref{sec:Preliminaries}. Based on this, we propose our model specification relaxation strategy in detail based on infinite-horizon optimal control in~\Cref{sec:proposedApproach}. After that, the experimental results are given in~\Cref{sec:exPerimentalResults} to demonstrate the efficacy of the proposed model specification relaxation strategy. Finally, conclusions and future research directions are summarized in~\Cref{sec:Conclusions}.

\section{Related Works}\label{sec:RelatedWorks}
In this section, we review the development of PLVMs to highlight the technical gaps and motivations underpinning this paper. Over the past decades, PLVMs have been widely employed in automation community for applications such as industrial process monitoring and inferential sensors, largely due to their ability to handle measurement noise and data uncertainty~\cite{kong2022latent, yu2022generalized,raveendran2018process}. Early PLVMs were predominantly based on a linear projection of the data generative process, enabling straightforward latent variable inference via matrix inversion techniques~\cite{bishop2006pattern,murphy2022probabilistic}. However, the linear assumption inherently limited the expressiveness of these models, especially for complex downstream tasks. To overcome this limitation, researchers have progressively incorporated advanced neural architectures into PLVMs, such as multi-layer perceptrons~\cite{kingma2013auto}, recurrent neural networks~\cite{9638604}, and Transformers~\cite{tang2021probabilistic}. These advancements significantly enhanced the representational capacity of PLVMs by leveraging inductive biases tailored to specific datasets, allowing for more flexible modeling of non-linear data structures.

Despite these progresses, challenges associated with regularization remain a critical bottleneck. Traditional approaches typically require constraining the latent variables to a predefined normalized distribution family to reduce computational complexity and facilitate the inclusion of regularization terms~\cite{knoblauch2022optimization}. While this constraint simplifies the optimization process, it often compromises model flexibility and limits approximation accuracy, thereby impairing performance on downstream tasks.

Recent developments in stochastic differential equation theory~\cite{10547229,7549018,chen2019particle} and transportation theory~\cite{7160692,7160709} have inspired the emergence of diffusion models~\cite{yang2023diffusion, songSGM}, which take an alternative approach to regularization. Rather than focusing solely on the latent variables, these models regularize the generative process mapping latent variables to observational data, thereby relaxing the regularization domain to the \emph{path space}~\cite{zhang2021path}.
% Instead of focusing on the latent variables, these models regularize the generative process from latent variables to observational data, expanding the regularization domain to the \emph{path space}. 
This paradigm shift has achieved remarkable success in diverse applications, including image generation~\cite{ho2020denoising} and data imputation~\cite{chen2023provably,tashiro2021csdi}. However, while regularizing the generative process has improved flexibility in modeling, its application to PLVMs remains underexplored. Specifically, the method of adapting this strategy to PLVMs and developing a novel, convergence-guaranteed EM algorithm for training PLVMs has not been well addressed. Building on the advancements introduced by diffusion models, we identify two significant technical gaps that this work aims to address in the context of PLVMs:

\begin{enumerate}[leftmargin=*]  
\item{\textbf{Relaxation of Latent Variable Specification:} Can a rigorous relaxation strategy be designed for PLVMs to improve their performance on downstream tasks?}
\item{\textbf{Implementation of Relaxation in PLVM Training:} Can a novel convergence-guaranteed EM algorithm be developed to incorporate the relaxation strategy into PLVM training effectively? }
\end{enumerate}

\section{Preliminaries}\label{sec:Preliminaries}
\subsection{Notations}\label{subsec:Notations}
The notations in this paper are briefly introduced as follows:
\begin{itemize}[leftmargin=*] 
\item{\textbf{Script Notations:} The calligraphic font is designed to denote the probability density function (PDF). For example, $\mathcal{Q}:\mathbb{R}^{\mathrm{D}}\rightarrow\mathbb{R}^{+}\cup\{0\}$ represents the PDF of the $\mathrm{D}$-dimensional random variable $z\in\mathbb{R}^{\mathrm{D}}$.  }
\item{\textbf{Lower Case:} The lower case is designed to denote the function that parameterizes the PDF. For instance, $p_{\theta}:\mathbb{R}^{\mathrm{D}}\rightarrow\mathbb{R}^{+}\cup\{0\}$ indicates the PDF $\mathcal{P}(z)$ is parameterized by the function $p$ with parameter (set) $\theta$. }
\item{\textbf{Support of Distributions:} All probability distributions discussed herein are supported on the $\mathrm{D}$-dimensional real number domain $\mathbb{R}^{\mathrm{D}}$.}
\item{\textbf{Time Index:} Finally, we adopt $t$ as time index.}
\end{itemize} 
\subsection{The Architecture and EM Algorithm of PLVMs }\label{subsec:PLVMandItsLearning}

\begin{figure}[htbp]
    \centering
    % \vspace{-0.3cm}
    % picture\IllustrationPLVM.pdf
    \includegraphics[width=0.50\columnwidth]{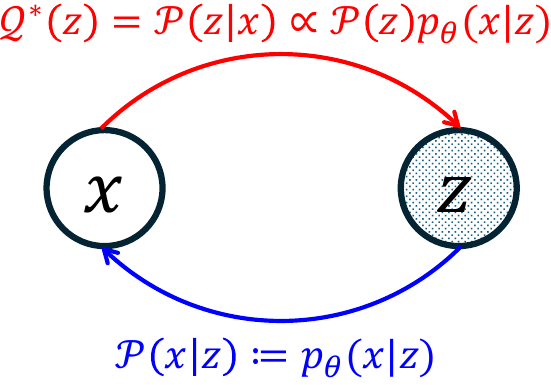}
    % \vspace{-0.2cm}     
    \caption{
    The illustration of PLVMs and their learning procedure.
    % The illustration of (a) AVI, (b) generative network (decoder) of NDPLVM for inferential sensor modeling, and (c) inference network (encoder) of NDPLVM for inferential sensor modeling. The red cross in (a) indicates that reversing $p_{\theta}(x\vert z)$ is impossible when model $p_{\theta}$ by neural network. The blue shaded part in (b) is the Markov blanket of node $z_1$. 
    }\label{fig:PLVMStructure}
   % \vspace{-0.5cm}     
\end{figure}

The model architecture for PLVMs is given in~\Cref{fig:PLVMStructure}, where $x\in\mathbb{R}^{\mathrm{D}_{\rm{obs}} \times 1}$ is the observational data (`$\rm{obs}$' is the abbreviation of `observational'), and $z\in\mathbb{R}^{\mathrm{D}_{\rm{LV}}\times 1}$ is the latent variable (`$\rm{LV}$' is the abbreviation of `latent variable'). The data generative procedure for PLVM can be delineated by the blue arrow, where data $x$ is predicted by function $p_{\theta}(x|z)$ with parameter $\theta$ and input $z$, and the conditional likelihood $\mathcal{P}(x|z)$ is modeled by function $p_{\theta}(x|z)$ as $\mathcal{P}(x|z)\coloneqq p_\theta(x|z)$. Based on this, denote the prior distribution of latent variable as $\mathcal{P}(z)$; the learning objective for PLVM is given as follows:
\begin{equation}\label{eq:ELBODefinition}
    \begin{aligned}
        \log{\mathcal{P}(x)}
       % =& \log{\int{\mathcal{P}(x,z)\mathrm{d}z}}\\
      %  =& \log{\int{\mathcal{P}(x,z)\times \frac{\mathcal{Q}(z)}{\mathcal{Q}(z)}\mathrm{d}z}} \\
      %  =& \log{\mathbb{E}_{\mathcal{Q}(z)}\left[\frac{\mathcal{P}(x,z)}{\mathcal{Q}(z)}\right]} \\
       % \overset{\text{(i)}}{\ge} & \mathbb{E}_{\mathcal{Q}(z)}\left[\log{\frac{\mathcal{P}(x,z)}{\mathcal{Q}(z)}}\right] \\
      %  = & \mathbb{E}_{\mathcal{Q}(z)}\left[\log{\mathcal{P}(x|z)}\right]-\mathbb{D}_{\rm{KL}}\left[\mathcal{Q}(z)\Vert \mathcal{P}(z)\right]\\
     \overset{\text{(i)}}{\ge}  & \underbrace{\mathbb{E}_{\mathcal{Q}(z)}[\log{p_{\theta}(x|z)}] - \mathbb{D}_{\rm{KL}}\left[\mathcal{Q}(z)\Vert \mathcal{P}(z)\right]}_{\coloneqq \mathcal{L}\left(\theta,\mathcal{Q}(z)\right)},
        \end{aligned}
\end{equation}
where $\mathcal{Q}(z)$ is referred to as `approximation distribution', `(i)' is based on the celebrated Jensen's inequality, non-negative term $ \mathbb{D}_{\rm{KL}}\left[\mathcal{Q}(z)\Vert \mathcal{P}(z)\right]$ is the Kullback-Leiber divergence between $\mathcal{Q}(z)$ and $ \mathcal{P}(z)$, $\mathbb{E}$ is the expectation operator, and the last line is named Evidence Lower BOund (ELBO)~\cite{8409977}. 

The learning algorithm for PLVM is EM algorithm, which is realized by executing the following steps named `Expectation-step' (E-Step) and `Maximization-step' (M-step) iteratively based on the maximum-likelihood estimation principle:
\begin{equation}\label{eq:VBEMAlgorithm}
\begin{cases}
&\text{E-step:}~ \mathcal{Q}^{\tau+1}(z)=\mathop{\arg\max}_{\mathcal{Q}(z)}\mathcal{L}\left(\theta,\mathcal{Q}(z)\right)|_{\theta=\theta^{\tau}}\\
&\text{M-step:}~ \theta^{\tau+1}=\mathop{\arg\max}_{\theta}\mathcal{L}\left(\theta,\mathcal{Q}(z)\right)|_{\mathcal{Q}(z)=\mathcal{Q}^{\tau+1}(z)}
% &\text{M-step:}~ \theta^{t+1}=\mathop{\arg\max}_{\theta}\mathbb{E}_{\mathcal{Q}(z)}[\log{p_{\theta}(x|z)}]|_{\mathcal{Q}(z)=\mathcal{Q}^{t+1}(z)}
\end{cases},
\end{equation}
where the superscript $\tau$ indicates the iterative time. 

% 这里的$t$要换成$\tau$
Notably, since $\mathbb{E}_{\mathcal{Q}(z)}\left[\log{\mathcal{P}(x)}\right]$ is a constant, E-step can be reformulated as follows:
\begin{equation}\label{eq:Q_tELBOExpression}
\begin{aligned}
   & \mathcal{Q}^{\tau+1}(z)=\mathop{\arg\max}_{\theta=\theta^{\tau}}\mathcal{L}\left(\theta,\mathcal{Q}(z)\right)\Rightarrow\\  &  \mathcal{Q}^{\tau+1}(z)=\mathop{\arg\max}_{\theta}\underbrace{\mathcal{L}\left(\theta,\mathcal{Q}(z)\right) - \mathbb{E}_{\mathcal{Q}(z)}\left[\log{\mathcal{P}(x)}\right]}_{=-\mathbb{D}_{\rm{KL}}\left[\mathcal{Q}(z)\Vert \mathcal{P}(z|x)\right]\le 0},
\end{aligned}
\end{equation}
where the optimal variational distribution $\mathcal{Q}(z)$ denoted by $\mathcal{Q}^*(z)$ is obtained at posterior distribution $\mathcal{P}(z|x) $. According to the Bayesian formula $\mathcal{P}(z|x)\propto p_{\theta}(x|z)\mathcal{P}(z)$ as the red arrow in~\Cref{fig:PLVMStructure} shows. Furthermore,  for M-step, $\mathbb{D}_{\rm{KL}}\left[\mathcal{Q}(z)\Vert \mathcal{P}(z)\right]$ is a constant. Based on the abovementioned information,~\Cref{eq:VBEMAlgorithm} can be further reformulated as follows:
\begin{equation}\label{eq:EMFunctionIteration}
\begin{cases}
&\text{E-step:}~ \mathcal{Q}^{\tau+1}(z)=\mathcal{P}(z|x)\propto p_{\theta}(x|z)\mathcal{P}(z) |_{\theta=\theta^{\tau}}\\
% &\text{M-step:}~ \theta^{t+1}=\mathop{\arg\max}_{\mathcal{Q}(z)=\mathcal{Q}(z^{t+1})}\mathcal{L}\left(\theta,\mathcal{Q}(z)\right)
&\text{M-step:}~ \theta^{\tau+1}=\mathop{\arg\max}_{\theta}\mathbb{E}_{\mathcal{Q}(z)}[\log{p_{\theta}(x|z)}]|_{\mathcal{Q}(z)=\mathcal{Q}^{\tau+1}(z)}
\end{cases}.
\end{equation}

\subsection{Continuity Equation and Its Weak Solution}\label{subsec:continuityEquationWeakSolution}
Most PLVM frameworks focus on optimizing a probability density $\mathcal{Q}(z)$ to approximate the posterior distribution $\mathcal{P}(z|x)$. However, a powerful alternative is to view $\mathcal{Q}(z)$ as a time-evolving probability density, which transitions smoothly from an initial state $\mathcal{Q}_0(z)$ to a specified target state $\mathcal{Q}_{\mathrm{T}}(z)$ gradually. The mathematical foundation of this approach is the celebrated \emph{continuity equation}, which describes how the probability density $\mathcal{Q}_t(z)$ evolves under the influence of a perturbation direction $\phi:\mathbb{R}^{\mathrm{D}_{\rm{LV}}}\to\mathbb{R}^{\mathrm{D}_{\rm{LV}}}$:  
\begin{equation}\label{eq:continuityEquation}  
    \frac{\partial \mathcal{Q}_t(z)}{\partial t} + \nabla_z\cdot [\phi(z)\, \mathcal{Q}_t(z)] = 0,  
\end{equation}  
where $ \frac{\partial \mathcal{Q}_t(z)}{\partial t}$ is the \emph{Eulerian derivative}. By applying the chain rule,~\Cref{eq:continuityEquation} can be equivalently reformulated in terms of the \emph{Lagrangian derivative} as:  
\begin{equation}\label{eq:lagrangianQ}  
     \frac{\mathrm{d} \mathcal{Q}_t(z)}{\mathrm{d} t} = - \mathcal{Q}_t(z)\nabla_z\cdot \phi(z),  
\end{equation}  
where  $\frac{\mathrm{d}\mathcal{Q}_t(z)}{\mathrm{d}t}= \frac{\partial \mathcal{Q}_{t}(z)}{\partial t} + \left[\nabla_z{\mathcal{Q}_{t}(z)} \right]^\top\left[\phi(z)\right] $ relates the Lagrangian and Eulerian derivatives.
% \begin{equation}  
%     \frac{\mathrm{d}\mathcal{Q}_t(z)}{\mathrm{d}t}= \frac{\partial \mathcal{Q}_{t}(z)}{\partial t} + \left[\nabla_z{\mathcal{Q}_{t}(z)} \right]^\top\left[\phi(z)\right]  
% \end{equation}  

The continuity equation~\Cref{eq:continuityEquation} is not always solvable in the classical sense~\cite{evans2022partial}, especially when the density $\mathcal{Q}_t(z)$ is not smooth, as in the case of particle-based measures. Therefore, we rely on the concept of a weak solution. A time-dependent measure $\mathcal{Q}_t(z)$ is defined as a weak solution to~\Cref{eq:continuityEquation} if, for every test function $f(z)$ in the compact support function space $\mathscr{C}_c^\infty(\mathbb{R}^{\mathrm{D}_{\rm{LV}}})$, it satisfies the following integral equation:
\begin{equation} \label{eq:weak_form_continuity}
    \frac{\mathrm{d}}{\mathrm{d}t} \int f(z) \, \mathcal{Q}_t(z)\mathrm{d}z = \int \phi^\top(z)\nabla_z f(z) \mathcal{Q}_t(z)\mathrm{d}z .
\end{equation}
% This formulation transfers the derivative from the potentially non-smooth measure $\mathcal{Q}_t$ onto the smooth test function $f$ via integration by parts.
% A key result from transport theory is that such a weak solution can be constructed numerically. 
Based on this, if a collection of particles $\{z_{i,t}\}_{i=1}^\mathrm{M}$ evolves according to the ODE:
\begin{equation}\label{eq:particleflow}
    \frac{\mathrm{d} z_{i,t}}{\mathrm{d} t} = \phi(z_{i,t})
\end{equation}
with initial positions $\{z_{i,0}\}_{i=1}^\mathrm{M}$ sampled from an initial distribution $\mathcal{Q}_0(z)$ , the resulting empirical measure at time $t$, $\mathscr{Q}_t(z) = \frac{1}{\mathrm{M}} \sum_{i=1}^\mathrm{M} \updelta(z - z_{i,t})$ constitutes a weak solution to the continuity equation~\Cref{eq:continuityEquation}, in the sense that it satisfies the weak form~\Cref{eq:weak_form_continuity}. Here, the upright Greek letter $\updelta(\cdot)$ denotes the Dirac delta measure. This principle provides a rigorous bridge between the continuous PDE formulation and a practical, particle-based numerical implementation. It allows us to simulate a complex distributional flow by solving a set of simple ODEs. A detailed demonstration that the empirical measure given by $ \mathscr{Q}_t(z)$ satisfies the weak form defined in \Cref{eq:weak_form_continuity} is provided in Section S.I of Supplementary Material.

\section{Proposed Approach}\label{sec:proposedApproach}

% this paper adopts the script notation to denote probability density functions (PDFs). For instance, $\mathcal{Q}(z):\mathbb{R}^{\mathrm{D}}\rightarrow\mathbb{R}^{+}$ represents the PDF of the random variable $z$, where $z$ belongs to a D-dimensional real space, $\mathbb{R}^{\mathrm{D}}$. Besides, all probability distributions discussed herein are supported on the real number domain $\mathbb{R}$. Furthermore, lowercase letters are used to denote PDFs that are parameterized by function. For example, $\psi_\varphi(z):\mathbb{R}^{\mathrm{D}}\rightarrow\mathbb{R}^{+}$ signifies that the PDF of $z$, residing in $\mathbb{R}^{\mathrm{D}}$, is parameterized by function $q$ with parameters denoted by $\varphi$. Additionally, during the derivations, the latent variable is represented by $z \in \mathbb{R}^{\mathrm{D}_{\text{LV}}}$, where $\mathrm{D}_{\text{LV}}$ indicates the dimension of the latent space. Similarly, process variables/covariates are denoted by $x \in \mathbb{R}^{\mathrm{D}_{\text{PV}}}$, where $\mathrm{D}_{\text{PV}}$ indicates the dimension of the process variable. Based on this

% Based on this, the ELBO Based on this, the variational distribution $\mathcal{Q}(z)$ can be directly,  variational distribution is always specified into a predefined function family $\mathbb{F}$, 

\subsection{Relaxation of LV Specification via Optimal Control}\label{subsec:LVSpecificationRelaxation}
\subsubsection{Model Specification for PLVM Learning}\label{subsec:implicitLVSpec}
% \subsubsection{Background knowledge for PLVM}
% \subsubsection{}\label{subsubsec:implicitLVSpec} 
Based on~\Cref{eq:EMFunctionIteration}, in the classic EM algorithm, the E-step is simplified by choosing a prior, $\mathcal{P}(z)$, that is conjugate to the likelihood, $p_{\theta}(x|z)$. This ensures that the true posterior, $\mathcal{P}(z|x)$, and its variational approximation, $\mathcal{Q}(z)$, remain in the same simple family of distributions (e.g., Gaussian). While computationally convenient, this conjugate prior constraint severely limits the flexibility of the variational distribution. By forcing $\mathcal{Q}(z)$ to belong to a predefined family, $\mathbb{F}$—such as the Gaussian family $\mathcal{N}(\mu, \Sigma)$—the model may fail to capture complex, multi-modal posteriors, as illustrated in Figure~\ref{subfig:specifiedApproximation}. This gap inevitably restricts the model's performance on downstream tasks, motivating the need for a more expressive approximation method.
\iffalse
Based on~\Cref{eq:EMFunctionIteration}, to streamline the computational procedure of the E-step, the prior distribution $\mathcal{P}(z)$ and evidence function $p_{\theta}(x|z)$ must be judiciously tailored to ensure that the posterior distribution $\mathcal{P}(z|x)$ remains within the same family as prior distribution $\mathcal{P}(z)$. Although this approach of employing the `conjugate prior' simplifies the computation of the variational distribution $\mathcal{Q}(z)$, it may inadvertently curtail flexibility of $\mathcal{Q}(z)$, which further restricts model performance on downstream tasks. 

Specifically, this methodology constraints $\mathcal{Q}(z)$ to a predefined family of normalized distributions, $\mathbb{F}$, such as the Gaussian distribution family defined by $\mathcal{Q}(z)\in\mathbb{F} \coloneqq \{\mathcal{N}(\mu, \Sigma) | \mu \in \mathbb{R}^{\mathrm{D}_{\rm{LV}}}, \Sigma \in \mathbb{R}^{\mathrm{D}_{\rm{LV}} \times \mathrm{D}_{\rm{LV}}}, \Sigma \succ 0\}$. This restriction limits the flexibility of $\mathcal{Q}(z)$, potentially compromising the overall model performance due to the lack of flexibility. The illustration can be given in~\Cref{subfig:specifiedApproximation}. When the posterior distribution $\mathcal{P}(z|x)$ is a mixture of Gaussian distribution, inferring the approximation $\mathcal{Q}(z)$ from $\mathbb{F}$ cannot cover the true posterior, thereby limiting the model performance inevitably.
\fi
\subsubsection{Inference from the Perspective of Differential Equation Simulation}\label{subsubsec:ODESimulationVariationalInference}
\begin{figure}[!h]
    % \vspace{-0.5cm}
\centering
     % figures\cut_sampled_results.pdf 
     % picture\cut_plot
     % D:\PycharmProjects\SteinVariational\ParVI\PrecondVIOC\exper_pfg\plot_baseline\cut_plot\cut_gmm_scatter_plot_1.pdf
\subfigure[$\mathcal{Q}(z)\in\mathbb{F}, \mathbb{F}=\{\mathcal{N}(\mu, \Sigma) | \mu \in \mathbb{R}^{\mathrm{D}_{\rm{LV}}}, \Sigma \in \mathbb{R}^{\mathrm{D}_{\rm{LV}} \times \mathrm{D}_{\rm{LV}}}, \Sigma \succ 0\}$]{\includegraphics[width=0.485\linewidth]{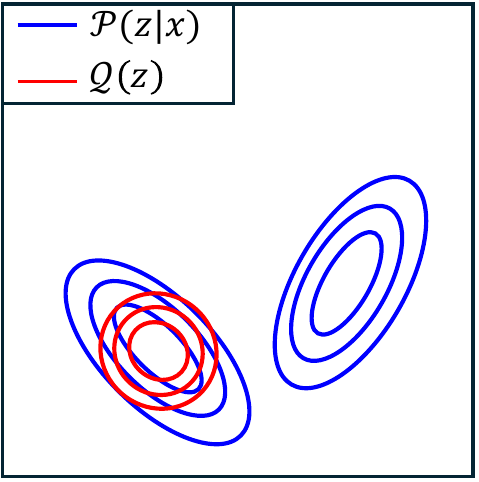}\label{subfig:specifiedApproximation}}
\subfigure[$\mathcal{Q}(z)=\frac{1}{\mathrm{M}}\sum_{i=1}^{\mathrm{M}}{\updelta (z-z_i)}$ ]{\includegraphics[width=0.485\linewidth]{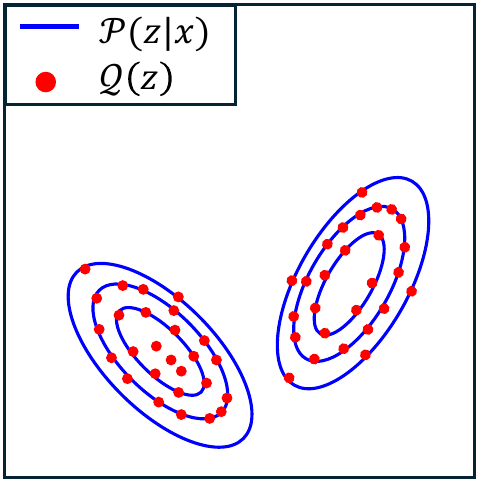}\label{subfig:particleApproximationFlow}}

\caption{The illustration for approximating $\mathcal{P}(z|x)$ with $\mathcal{Q}(z)$. 
}\label{fig:approximateIllustration}
\end{figure}

To overcome the limitations of predefined parametric families, we propose to \emph{quantize} the probability measure $\mathcal{Q}(z)$ by representing it with a finite set of $\mathrm{M}$ uniformly weighted particles $\{ z_i \}_{i=1}^{\mathrm{M}}$. This leads to the following approximation equation:
\begin{equation}\label{eq:QAsParticleMeasure}
    \mathcal{Q}(z) \approx \mathscr{Q}(z) = \frac{1}{\mathrm{M}} \sum_{i=1}^{\mathrm{M}} \updelta(z - z_{i}).
\end{equation}
As illustrated in~\Cref{subfig:particleApproximationFlow}, the shape of $\mathcal{Q}(z)$ can be flexibly adapted by adjusting the locations of the particles $\{ z_i \}_{i=1}^{\mathrm{M}}$. This quantization process enables our variational family to approximate a wide range of complex distributions far beyond those permitted by traditional parametric assumptions.

\begin{figure}[htbp]
    \centering
    % \vspace{-0.3cm}
    % picture\IllustrationPLVM.pdf
    \includegraphics[width=1.0\columnwidth]{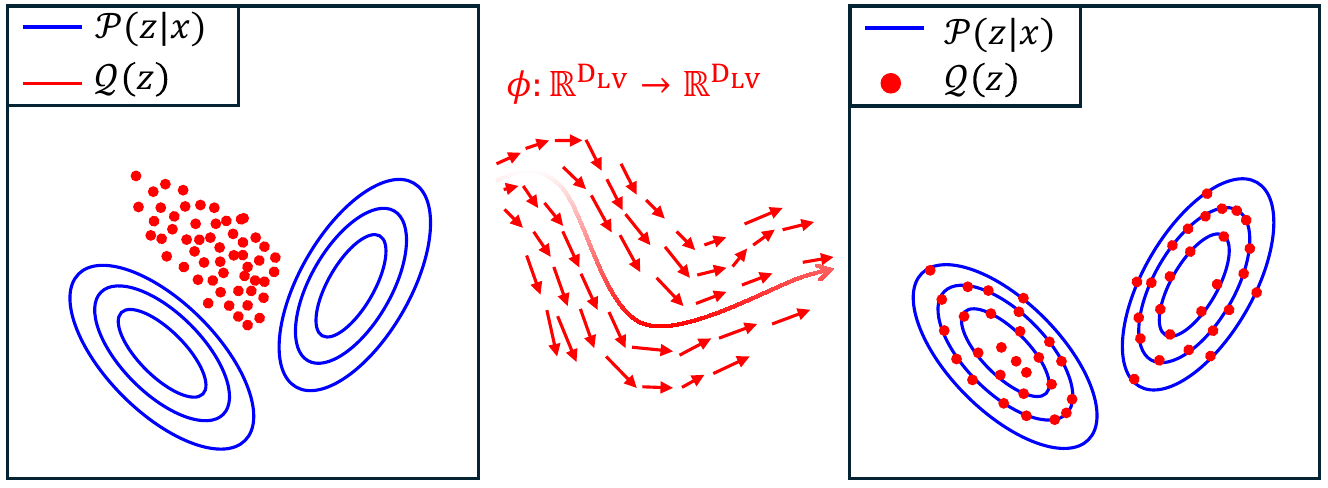}
    % \vspace{-0.2cm}     
    \caption{
    The illustration for the evolution of $\{ z_i \}_{i=1}^{\mathrm{M}}$ perturbed by $\phi(z)$.
    % The illustration of (a) AVI, (b) generative network (decoder) of NDPLVM for inferential sensor modeling, and (c) inference network (encoder) of NDPLVM for inferential sensor modeling. The red cross in (a) indicates that reversing $p_{\theta}(x\vert z)$ is impossible when model $p_{\theta}$ by neural network. The blue shaded part in (b) is the Markov blanket of node $z_1$. 
    }\label{fig:evolutionQzByPhiZ}
   % \vspace{-0.5cm}     
\end{figure}

While representing $\mathcal{Q}(z)$ using $\{ z_i \}_{i=1}^{\mathrm{M}}$ enhances model flexibility and improves approximation accuracy, the initial placement of $\{ z_i \}_{i=1}^{\mathrm{M}}$ may not guarantee such accuracy as the left part of~\Cref{fig:evolutionQzByPhiZ} shows. 
To address this issue, we introduce the ODE defined in~\Cref{eq:particleflow}, which progressively perturbs the particles $\{ z_i \}_{i=1}^{\mathrm{M}}$ by $\phi(z)$ over time $t$. This approach enables us to dynamically control the evolution of the empirical measure $\mathscr{Q}_t(z)$, thereby gradually optimizing the probability distribution $\mathcal{Q}_t(z)$ towards an accurate approximation of the true posterior $\mathcal{P}(z|x)$ by the final time $\mathrm{T}$, as illustrated in~\Cref{fig:evolutionQzByPhiZ}.
\begin{remark}
The particle evolution induced by the ODE in~\Cref{eq:particleflow} is not merely a heuristic. As established in~\Cref{subsec:continuityEquationWeakSolution}, the empirical measure constructed from the set $\{z_{i,t}\}_{i=1}^{\mathrm{M}}$ constitutes a \emph{weak solution} to the underlying continuity equation. Thus, by designing the perturbation direction $\phi(z)$, we directly steer the evolution of the probability density $\mathcal{Q}_t(z)$ via its governing PDE, ensuring both flexibility and theoretical rigorousness in our approximation strategy.
\end{remark}

\begin{figure}[htbp]
    \centering
    % \vspace{-0.3cm}
    % picture\IllustrationPLVM.pdf
    \includegraphics[width=0.95\columnwidth]{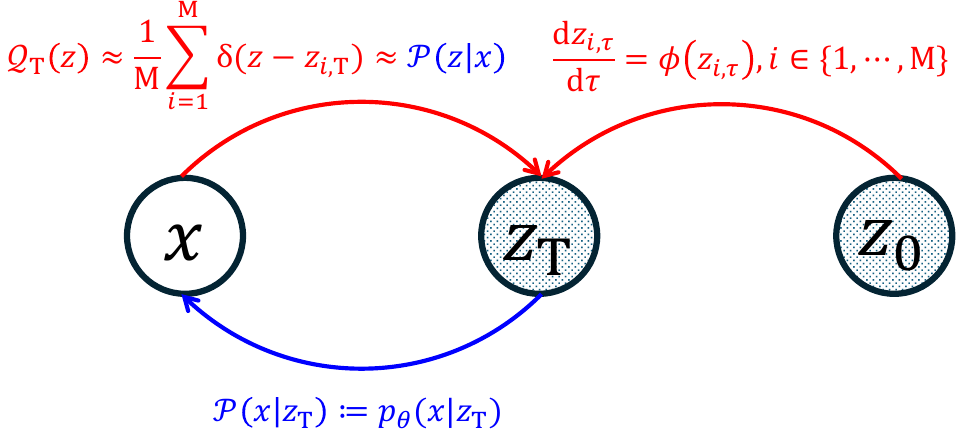}
    % \vspace{-0.2cm}     
    \caption{
    The illustration of PLVM, where $\{ z_i \}_{i=1}^{\mathrm{M}}$ are perturbed by $\phi(z)$.
    % The illustration of (a) AVI, (b) generative network (decoder) of NDPLVM for inferential sensor modeling, and (c) inference network (encoder) of NDPLVM for inferential sensor modeling. The red cross in (a) indicates that reversing $p_{\theta}(x\vert z)$ is impossible when model $p_{\theta}$ by neural network. The blue shaded part in (b) is the Markov blanket of node $z_1$. 
    }\label{fig:PLVMStructureODE}
   % \vspace{-0.5cm}     
\end{figure}

Consequently, the structure of the PLVM illustrated in~\Cref{fig:PLVMStructure} can be reformulated as shown in~\Cref{fig:PLVMStructureODE}. However, it should be pointed out that we mainly focus on changing the location of $\{ z_i \}_{i=1}^{\mathrm{M}}$, but the approximation accuracy is determined by the scalar function $\mathcal{Q}(z)$. 
% On this foundation, as~\Cref{fig:PLVMStructureODE} shows, we can introduce the ODE given in~\Cref{eq:ProgressiveTransformationODE} to perturb sample $\{ z_i \}_{i=1}^{\mathrm{M}}$ over time $t$ gradually, and realize the optimization of $\mathcal{Q}(z)$ to realize the goal of inferring the posterior distribution $\mathcal{P}(z|x)$ when $t$ reaches the end time $\mathrm{T}$. To this end, we first introduce the following theorem to delineate the evolution of $\mathcal{Q}(z)$ along time $t$ when perturbing $z$ according to the ODE defined by~\Cref{eq:ProgressiveTransformationODE}:

\subsubsection{Optimal Control Problem Formulation}\label{subsubsec:OCProblemFormulation}
In~\Cref{subsubsec:ODESimulationVariationalInference}, we have reformulated the inference problem as an ODE simulation problem. However, the perturbation direction $\phi(z)$, which ultimately determines the shape of $\mathcal{Q}_t(z)$ has not yet been determined.
% that drives the evolution of ODE has not yet been determined. 
To address this, we introduce an optimal control framework to identify $\phi(z)$ in this part.

We begin by reformulating the final expression in~\Cref{eq:Q_tELBOExpression} as:
\begin{equation}
    \mathbb{D}_{\rm{KL}}\left[\mathcal{Q}(z)\Vert \mathcal{P}(z|x)\right] = \mathbb{E}_{\mathcal{Q}(z)}\left[\log{\mathcal{Q}(z)}\right] - \mathbb{E}_{\mathcal{Q}(z)}\left[\log{\mathcal{P}(z|x)}\right],
\end{equation}
where we define $ \mathbb{E}_{\mathcal{Q}(z)}\left[\log{\mathcal{Q}(z)}\right]$ as the `negative entropy term' and $ - \mathbb{E}_{\mathcal{Q}(z)}\left[\log{\mathcal{P}(z|x)}\right]$ as the `negative log-likelihood term'. Since the parameters of $\mathcal{P}(z|x)$ remain fixed during the optimization of $\mathcal{Q}(z)$, we represent $\mathcal{Q}_t(z)$ using the empirical measure $\mathscr{Q}_t(z)$ as defined in~\Cref{eq:QAsParticleMeasure}. Using this representation, the negative log-likelihood term can be approximated via the celebrated Monte Carlo method as:
% where we name $ \mathbb{E}_{\mathcal{Q}(z)}\left[\log{\mathcal{Q}(z)}\right]$ as entropy term and $ - \mathbb{E}_{\mathcal{Q}(z)}\left[\log{\mathcal{P}(z|x)}\right]$ as negative log-likelihood term. Note that, since the parameter of $\mathcal{P}(z|x)$ is fix during optimizing $\mathcal{Q}(z)$, we represent $\mathcal{Q}(z)$ by $\{ z_i \}_{i=1}^{\mathrm{M}}$ according to~\Cref{eq:QAsParticleFunction} and we can approximate $ - \mathbb{E}_{\mathcal{Q}(z)}\left[\log{\mathcal{P}(z|x)}\right]$ by the celebrated Monte Carlo approximation as follows: 
\begin{equation}\label{eq:MCIIDQzFunction}
     \mathbb{E}_{\mathcal{Q}(z)}\left[\log{\mathcal{P}(z|x)}\right]\approx \mathbb{E}_{\mathscr{Q}(z)}\left[\log{\mathcal{P}(z|x)}\right] =\frac{1}{\mathrm{M}}\sum_{i=1}^{\mathrm{M}}{\log{\mathcal{P}(z_i|x)}}. %\{ z_i \}_{i=1}^{\mathrm{M}}\sim\mathcal{Q}(z).
\end{equation}
Based on~\Cref{eq:MCIIDQzFunction}, we observe that optimizing the negative log-likelihood term $-\mathbb{E}_{\mathcal{Q}(z)}\left[\log{\mathcal{P}(x|z)}\right]$ in KL divergence can be realized by the gradient descent method~\cite{boyd2004convex}, which perturbs $\{ z_i \}_{i=1}^{\mathrm{M}}$ by the following equation:
\begin{equation}
    z_i\leftarrow z_i + \varepsilon \nabla_z{{\log{\mathcal{P}(z|x)}}}\vert_{z=z_i},~\text{for}~i=1,2,\dots,\mathrm{M},
\end{equation}
which can be regarded as the discretization of the following ODE by forward Euler's method~\cite{butcher2016numerical,ambrosio2005gradient}:
\begin{equation}
    \frac{\mathrm{d} z_{i}}{\mathrm{d}t}=\nabla_z{{\log{\mathcal{P}(z|x)}}}\vert_{z=z_{i}},~\text{for}~i=1,2,\dots,\mathrm{M}.
\end{equation}
Based on the above analysis, the perturbation $\phi(z)$ that progressively reduces the KL divergence $\mathbb{D}_{\rm{KL}}\left[\mathcal{Q}(z)\Vert \mathcal{P}(z|x)\right]$ can be expressed as:
\begin{equation}\label{eq:perturbationEquationResult}
    \phi(z) = \nabla_z\log{\mathcal{P}(z|x)} + u(z),
\end{equation}
where we optimize the negative log-likelihood term with the help of $\nabla_z\log{\mathcal{P}(z|x)}$, and \emph{introduce a `control policy' $u(z)$ to optimize the negative entropy term}. Notably, control policy $u(z)$ belongs to the path space $C\left([0,\mathrm{T}], \mathbb{R}^{\mathrm{D}_{\rm{LV}}}\right)$, i.e., $u(z) \in C\left([0,\mathrm{T}], \mathbb{R}^{\mathrm{D}_{\rm{LV}}}\right)$. Accordingly, the following optimal control problem is formulated for the evolution of $t$ up to time $\mathrm{T}$:  
\begin{align}
  &  \begin{aligned}\mathop{\arg\min}_{u(z)}  \indent &  \mathbb{D}_{\rm{KL}}\left[\mathcal{Q}_{\mathrm{T}}(z)\Vert \mathcal{P}(z|x)\right] +\frac{1}{2}\int_{0}^{\mathrm{T}}{u^\top(z)u(z)\mathrm{d}t}, 
\end{aligned} 
    \label{eq:OCLVMObjectiveDetermin}\\
&    \mathrm{s.t.} \indent \begin{cases} & \dfrac{\mathrm{d} \mathcal{Q}_t(z) }{\mathrm{d} t}=-\mathcal{Q}_t(z)\nabla_z\cdot \left[ \nabla_z\log{\mathcal{P}(z|x)} + u(z)\right] \\ 
& \mathcal{Q}_t(z)|_{t=0} =\mathcal{Q}_0(z)
\end{cases} \tag{\ref{eq:OCLVMObjectiveDetermin}{a}} \label{eq:OCLVMObjectiveDeterminConstraint},
\end{align}
Here, the term $\frac{1}{2}\int_{0}^{\mathrm{T}} u^\top(z) u(z) \mathrm{d}t$ is introduced to regularize the control policy $u(z)$, as there may be infinitely many possible $u(z)$ that minimize $ \mathbb{D}_{\rm{KL}}\left[\mathcal{Q}_{\mathrm{T}}(z)\Vert \mathcal{P}(z|x)\right]$. \emph{Notably, no explicit regularization is imposed on the state $z$ or the probability density function $\mathcal{Q}_t(z)$, because the state evolution is fully determined by the control policy $u(z)$ and the initial condition via the continuity equation. Regularizing $u(z)$ is thus sufficient and standard in optimal control formulations.} After solving the optimal control problem, we can therefore obtain the weak solution by simulating the following ODE directly:
\begin{equation}
 \begin{cases} & \dfrac{\mathrm{d} z_{i,t} }{\mathrm{d} t}= \nabla_z\log{\mathcal{P}(z|x)} + u(z)|_{z=z_{i,t}},\\ 
& z_{i,t}\overset{\mathrm{i.i.d.}}{\sim}\mathcal{Q}_0(z).
\end{cases} 
\end{equation}
% Notably, we introduce $\frac{1}{2}\int_{0}^{\mathrm{T}}{u^\top(z)u(z)\mathrm{d}t}$ to regularize the control policy since there might be infinity possible $u(z)$ can realize the minimization of KL divergence $ \mathbb{D}_{\rm{KL}}\left[\mathcal{Q}_{\mathrm{T}}(z)\Vert \mathcal{P}(z|x)\right]$. 

Several approaches can be employed to solve the optimal control problem defined in~\Cref{eq:OCLVMObjectiveDetermin,eq:OCLVMObjectiveDeterminConstraint}. For instance, a control policy of the form $u(z) = -\mathcal{K}(z) \widehat{\mathcal{Q}}_t(z)$, (where $\mathcal{K}(z)$ and $\widehat{\mathcal{Q}}_t(z)$ denote the control gain and an estimate of ${\mathcal{Q}}_t(z)$, respectively), can be designed~\cite{lewis2012optimal,9658150}. However, state estimation poses significant challenges since $\mathcal{Q}_t(z)$ is represented as by the empirical measure $\mathscr{Q}_t(z)$, as shown in~\Cref{eq:QAsParticleMeasure}, which makes the real-time estimation of the intractable $\mathcal{Q}_t(z)$ and the computation of the KL divergence term $\mathbb{D}_{\rm{KL}}\left[\mathcal{Q}_{t}(z) \Vert \mathcal{P}(z|x)\right]$ particularly difficult.
% Several methods can be used to solve the optimal control problem defined in~\Cref{eq:OCLVMObjectiveDetermin,eq:OCLVMObjectiveDeterminConstraint}. For example, a control policy of the form $u(z) = -\mathcal{K}(z) \widehat{\mathcal{Q}}(z)$ ($\mathcal{K}(z)$ and $\widehat{\mathcal{Q}}(z)$ are control gain and estimation of ${\mathcal{Q}}(z)$, respectively) can be designed. However, state estimation remains challenging since $\mathcal{Q}(z)$ is represented by $\{ z_i \}_{i=1}^{\mathrm{M}}$ as shown in~\Cref{eq:QAsParticleFunction}, making real-time estimation of the intractable $\mathcal{Q}(z)$ and computation of the KL divergence term $\mathbb{D}_{\rm{KL}}\left[\mathcal{Q}_{\mathrm{T}}(z) \Vert \mathcal{P}(z|x)\right]$ difficult.
%%%%%%%%%%%%%%%%%%%%%%%%%%%%%%
% There are several ways to compute an optimal control problem defined by~\Cref{eq:OCLVMObjectiveDetermin,eq:OCLVMObjectiveDeterminConstraint}. For instance, we can design $u(z)=-K(z)\mathcal{Q}(z)$, but the state estimation remains difficulty since $\mathcal{Q}(z)$ is represented by $\{ z_i \}_{i=1}^{\mathrm{M}}$ as per~\Cref{eq:QAsParticleFunction} shows, which results in problem about the real-time estimation of the intractable $\mathcal{Q}(z)$ and the computation of the KL divergence term $\mathbb{D}_{\rm{KL}}\left[\mathcal{Q}_{\mathrm{T}}(z)\Vert \mathcal{P}(z|x)\right]$. 

To simplify the procedure, we consider the infinite-time limit $\mathrm{T} \to \infty$, where $\lim_{\mathrm{T} \to \infty} \mathbb{D}_{\rm{KL}}\left[\mathcal{Q}_{\mathrm{T}}(z) \Vert \mathcal{P}(z|x)\right] = 0$ (see subsequent contents for the justification for this condition). Under this assumption, the objective for the finite-horizon optimal control problem is reformulated as the following objective for the infinite-horizon optimal control problem:
% To ease the solving procedure, we consider the infinity time limit $\mathrm{T}\to\infty$, where $\lim_{\mathrm{T}\to\infty}{\mathbb{D}_{\rm{KL}}\left[\mathcal{Q}_{\mathrm{T}}(z)\Vert \mathcal{P}(z|x)\right]=0}$ (the justification of this condition is proved in the latter content) and reformulate the objective for finite-horizon optimal control problem into the following objective for infinite-horizon optimal control problem: 

\begin{align}
  &  \begin{aligned}\mathop{\arg\min}_{u(z)}  \indent &  \frac{1}{2}\int_{0}^{\infty}{u^\top(z)u(z)\mathrm{d}t}, , 
\end{aligned} 
    \label{eq:InfOCLVMObjectiveDetermin}\\
&    \mathrm{s.t.} \indent \begin{cases} & \dfrac{\mathrm{d} \mathcal{Q}_t(z) }{\mathrm{d} t}=-\mathcal{Q}_t(z)\nabla_z\cdot \left[ \nabla_z\log{\mathcal{P}(z|x)} + u(z)\right] \\ 
& \mathcal{Q}_t(z)|_{t=0} =\mathcal{Q}_0(z)
\end{cases} \tag{\ref{eq:InfOCLVMObjectiveDetermin}{a}} \label{eq:OCLVMObjectiveDeterminConstraint2},
\end{align}
Up to now, we have formulated the optimal control problem for latent variable distribution $\mathcal{P}(z|x)$ inference. 
\begin{remark}
The optimal control problem defined in~\Cref{eq:InfOCLVMObjectiveDetermin,eq:OCLVMObjectiveDeterminConstraint2} reformulates the latent variable distribution inference task—where $\mathcal{Q}(z) \in \mathbb{F}$ is inferred—into an optimal control problem, with the control policy $u(z)$ residing in the \emph{infinite-horizon path space} $C\left([0,\infty), \mathbb{R}^{\mathrm{D}_{\rm{LV}}}\right)$, i.e., $u(z) \in C\left([0,\infty), \mathbb{R}^{\mathrm{D}_{\rm{LV}}}\right)$. 
\end{remark}
\begin{remark}
The dynamical formulation introduces greater flexibility compared to static variational approaches, as it decouples the final distribution $\mathcal{Q}_{T}(z)$ from the potentially rigid constraints imposed by the original model family $\mathbb{F}$. Rather than directly enforcing restrictive structural assumptions on $\mathcal{Q}_{T}(z)$, the intermediate distributions $\mathcal{Q}_t(z)$ can iteratively evolve toward the optimum. This `path regularization' distributes the burden of regularization throughout the entire evolution process, thus significantly relaxing the specification constraints and enabling more expressive approximations.
\end{remark}
\subsection{Optimal Control Problem Solving, Analysis, and Implementation}
Before solving and implementing the optimal control problem to obtain the perturbation for weak solution simulation, the following assumptions are introduced:
\begin{assumption}\label{assump:zeroBoundCond} 
The function $\phi(z)$ has compact support, $\phi\in\mathscr{C}_c^{\infty}(\mathbb{R}^{\mathrm{D}_{\rm{LV}}})$; that is, there exists a positive constant $\mathscr{R} > 0$ such that $\phi(z) = 0$ for all $\|z\|  > \mathscr{R} $.  
% The function $\phi(x)$ has compact support, $\phi\in\mathscr{C}_c(\mathbb{R}^{\mathrm{D}})$; that is, there exists $R > 0$ such that $\phi(x) = 0$ for all $\|x\| > R$.  
\end{assumption}  
\begin{assumption}\label{assump:boundedCond}  
The probability density function $\mathcal{Q}(z)$ vanishes at infinity:  $\lim_{\|z\| \to \infty}\mathcal{Q}(z) = 0$.  
\end{assumption}

\iffalse

\begin{assumption}\label{assump:zeroBoundCond}
The perturbation vanishes as $\Vert z\Vert$ approaches infinity, i.e., $\lim_{\Vert z \Vert \to\infty}{\phi(z)}=0$.
\end{assumption}
\begin{assumption}\label{assump:boundedCond}
The density functions $\mathcal{Q}(z)$ and $\mathcal{P}(z\vert x)$ are bounded.
\end{assumption}

\fi
Based on~\Cref{assump:zeroBoundCond,assump:boundedCond}, we can introduce the following theorem that is concerned with $\phi(z)$ and $\mathcal{Q}(z)$:
\begin{theorem}\label{thm:equalZeroTheorem}
When~\Cref{assump:zeroBoundCond,assump:boundedCond} are satisfied, the following equation sets up:
\begin{equation}\label{eq:zeroBoundConditionResult}
   % \int_{\mathbb{R}^{\mathrm{D}_{\rm{LV}}}}{\nabla\cdot{\phi(z)\mathcal{Q}(z)}\mathrm{d}z}=0.
   \int{\phi^\top(z)\nabla_z\mathcal{Q}(z)\mathrm{d}z} = -\int{\mathcal{Q}(z)\nabla_z\cdot\phi(z)\mathrm{d}z}.
\end{equation}
\end{theorem}

\subsubsection{Optimal Control Problem Solving}\label{subsubsec:OCProblemSolvingResult}
Building on~\Cref{thm:equalZeroTheorem}, we propose the following theorem to solve the optimal control problem defined by~\Cref{eq:InfOCLVMObjectiveDetermin,eq:OCLVMObjectiveDeterminConstraint2}:
\begin{theorem}
The solution to problem~\Cref{eq:InfOCLVMObjectiveDetermin,eq:OCLVMObjectiveDeterminConstraint2} is given by the following optimal control law:
 \begin{equation}\label{eq:optimalControlLaw}
    u(z) = - \nabla_z\log{\mathcal{Q}_t(z)}.
 \end{equation} 
\end{theorem}

% Up to now, we have obtained the solution to the optimal control problem. 
On this basis, the ODE that drives the $\mathcal{Q}_t(z)$ approaches to $\mathcal{P}(z\vert x)$ can be given as follows:
\begin{equation}\label{eq:FinalFullDeriviateEquation}
    \frac{\mathrm{d}\mathcal{Q}_t(z)}{\mathrm{d}t}=-\mathcal{Q}_t(z)\nabla_z\cdot \left[ \nabla_z\log{\mathcal{P}(z|x)} -  \nabla_z\log{\mathcal{Q}_t(z)}\right].
\end{equation}
Consequently, the corresponding weak solution represented by empirical measure $\mathscr{Q}_t(z)$ can be further rectified as:
\begin{equation}\label{eq:FinalParticleDeriviateEquation}
   \frac{\mathrm{d}z_{i,t}}{\mathrm{d}t} = \phi(z) = \nabla_z\log{\mathcal{P}(z|x)} -\nabla_z\log{\mathcal{Q}_t(z)}|_{z=z_{i,t}}.
\end{equation}
Based on this equation, we have the following remark:
\begin{remark}\label{remark:scoreFunctionApplicability}
It can be observed that the inference of the latent variable distribution $\mathcal{P}(z|x)$ by simulating the ODE defined in~\Cref{eq:perturbationEquationResult} merely requires the score function $\nabla_z \log{\mathcal{P}(z|x)}$, which significantly simplifies the implementation of the inference procedure. Specifically, the score function $\nabla_z \log{\mathcal{P}(z|x)}$ can be decomposed as follows:
    \begin{equation}
    \begin{aligned}
        \nabla_{z}\log{\mathcal{P}(z|x)}&={\nabla_z\log{\mathcal{P}(z)}} + \underbrace{\nabla_z\log{\mathcal{P}(x|z)}}_{=\nabla_z\log{p_{\theta}(x|z)}}-\underbrace{\nabla_z\log{\mathcal{P}(x)}}_{=0}\\
        & = \nabla_z\log{\mathcal{P}(z)} + \nabla_z\log{p_{\theta}(x|z)},
    \end{aligned}
    \end{equation}
    where $\nabla_z\log{\mathcal{P}(z)}$ can be given analytically, and $\nabla_z\log{p_{\theta}(x|z)}$ can be efficiently computed using automatic differentiation frameworks such as PyTorch~\cite{TorchNips} and JAX~\cite{bradbury2021jax}. Notably, this decomposition naturally avoids the computation of the intractable \emph{logarithm of the normalized density function}, $\log{\mathcal{P}(x)}$, further simplifying the inference process.
    \end{remark}

\subsubsection{Equilibrium State Analysis}\label{subsubsec:equilibriumStateAnalysis}
Notably, in~\Cref{subsubsec:OCProblemFormulation,subsubsec:OCProblemSolvingResult}, we assume that $$\lim_{\mathrm{T}\to\infty}{\mathbb{D}_{\rm{KL}}\left[\mathcal{Q}_{\mathrm{T}}(z)\Vert \mathcal{P}(z|x)\right]=0}$$. It is crucial to justify this assumption. Hence, we propose the following theorem:
\begin{theorem}%[IV.2]
As $t \to \infty$,~\Cref{eq:FinalFullDeriviateEquation} approaches an equilibrium state, indicating the stabilization of the system. In this context, for the perturbation direction defined as $$\phi(z) = \nabla_z \log{\mathcal{P}(z|x)} - \nabla_z \log{\mathcal{Q}_t(z)},$$ the distribution $\mathcal{Q}_{t}(z)$ evolves according to~\Cref{eq:FinalFullDeriviateEquation} and asymptotically converges to the conditional probability distribution:
% As $t \to \infty$,~\Cref{eq:FinalFullDeriviateEquation} approaches an equilibrium state, reflecting the system's stabilization. In this context, for the perturbation direction defined by $\phi(z) = \nabla_z\log{\mathcal{P}(z|x)}- \nabla_z\log{\mathcal{Q}_t(z)}$, it is observed that the distribution $\mathcal{Q}_\mathrm{T}(z)$ evolves according to~\Cref{eq:FinalFullDeriviateEquation} and asymptotically converges to the conditional probability distribution as follows:
\begin{equation}
    \lim_{\mathrm{T}\to\infty}{\mathbb{D}_{\rm{KL}}\left[\mathcal{Q}_{\mathrm{T}}(z)\Vert \mathcal{P}(z|x)\right]=0}.
\end{equation}
\end{theorem}
\noindent Thus far, we have demonstrated that the equilibrium point of $\mathcal{Q}_t(z)$, as induced by~\Cref{eq:FinalFullDeriviateEquation}, converges to $\mathcal{P}(z|x)$, which further validates the condition $\lim_{\mathrm{T} \to \infty} \mathbb{D}_{\rm{KL}}\left[\mathcal{Q}_{\mathrm{T}}(z) \Vert \mathcal{P}(z|x)\right] = 0$ made in~\Cref{subsubsec:OCProblemFormulation,subsubsec:OCProblemSolvingResult}.
% So far, we have proved the equilibrium point of $\mathcal{Q}_t(z)$ induced by~\Cref{eq:FinalFullDeriviateEquation}, which converges to the $\mathcal{P}(z|x)$, which further proves the justification of: $\lim_{\mathrm{T}\to\infty}{\mathbb{D}_{\rm{KL}}\left[\mathcal{Q}_{\mathrm{T}}(z)\Vert \mathcal{P}(z|x)\right]=0}$ in~\Cref{subsubsec:OCProblemFormulation,subsubsec:OCProblemSolvingResult}.

\subsubsection{Ansatz for Weak Solution Implementation}\label{subsubsec:OCProblemImplementation}
In this part, the implementation of the weak solution via computer language is further discussed. In the beginning, we should realize the major difficulty of simulating~\Cref{eq:FinalParticleDeriviateEquation} to demonstrate the necessity of this part: Perturbing $\{ z_i \}_{i=1}^{\mathrm{M}}$ with $\phi(z)$ by simulating~\Cref{eq:FinalParticleDeriviateEquation} requires the real-time-estimation of $\nabla_z\log{\mathcal{Q}_t(z)}$, which constitutes the control policy $u(z)$ according to~\Cref{eq:optimalControlLaw}. However, $\mathcal{Q}_t(z)$ is represented by the empirical measure $\mathscr{Q}_t(z)$, which cannot be analytically solved according to reference~\cite{evans2022partial}. Consequently, how to sidestep explicitly estimating the $\mathcal{Q}_t(z)$ is of great necessity for the weak solution implementation. 

To address this issue, we introduce the ansatz\footnote{The term ``ansatz'' is borrowed from quantum mechanics~\cite{pfau2020ferminet}, where it refers to a trial or approximate form for the many-body wave function.} $\psi:\mathbb{R}^{\mathrm{D}_{\rm{LV}}}\to \mathbb{R}^{\mathrm{D}_{\rm{LV}}}$ to approximate the optimal control policy $\nabla_z \log{\mathcal{Q}_t}(z)$, and perturb $\{ z_i \}_{i=1}^{\mathrm{M}}$ according to the following equation:
\begin{equation}\label{eq:objective4VelocityFieldApprox}
   \dfrac{ \mathrm{d} z_{i,t}}{\mathrm{d}t}=\left[\nabla_{z}{\log{\mathcal{P}(z|x)}} +   \psi(z)\right]\vert_{z=z_{i,t}}.%~\text{for}~i=1,2,\dots,\mathrm{M}.
   % \dfrac{\mathrm{d}z}{\mathrm{d}t} = \psi_\varphi(z),
\end{equation}

To ensure that the ansatz $\psi(z)$ closely approximates the optimal control policy $u(z)$ in terms of the `perturbation direction', we formulate the following optimization objective for $\psi(z)$ based on inner product similarity:
% \begin{equation}\label{eq:ansatzOptimizationFunction}
%     \mathop{\arg\max}_{\psi(x)}\int{\mathcal{Q}_t(z) \left\Vert \left[\nabla_{z}{\log{\mathcal{P}(z|x)}} +   \psi(z)\right] - \left[\nabla_{z}{\log{\mathcal{P}(z|x)}} -\nabla_{z}{\log{\mathcal{Q}_t(x)}}\right] \right\Vert^2_2 \mathrm{d}z}
% \end{equation}
\begin{align}\label{eq:objForQVarPhiZ}
    \mathop{\arg\max}_{\psi(z)} \int \mathcal{Q}_t(z) \, &\left[\nabla_z{\log{\mathcal{P}(z|x)}} - \nabla_z{\log{\mathcal{Q}_t(z)}} \right]^\top \nonumber \\
    &\quad  \times \left[ \nabla_z{\log{\mathcal{P}(z|x)}} +\psi(z)\right] \, \mathrm{d}z.
\end{align}
However,~\Cref{eq:objForQVarPhiZ} remains an intractable term $ \nabla_{z}{\log{\mathcal{Q}_t(x)}}$. To alleviate this issue, we propose the following theorem to convert~\Cref{eq:objForQVarPhiZ} into an equivalent form:
%Since the key to perturbing $\{ z_i \}_{i=1}^{\mathrm{M}}$ is the `perturbation direction', we design the following objective function based on the inner product similarity to evaluate the approximation accuracy of the ansatz $\psi(z)$:
% \begin{equation}\label{eq:objForQVarPhiZ}
% \begin{aligned}
% \left\langle\nabla_z{\log{\mathcal{P}(z|x)}} - \nabla_z{\log{\mathcal{Q}_t(z)}}, \nabla_z{\log{\mathcal{P}(z|x)}} +\psi(z)\right\rangle_{\mathcal{Q}_t(z)},
% \end{aligned}
% \end{equation}
% On this basis, we can formulate the learning objective to optimize the ansatz $\psi(z)$:
\begin{theorem}\label{thm:rkhsVelocityExpression}
% Given the following conditions: 1) the ansatz vanishes as $\Vert z\Vert$ approaches infinity, i.e., $\lim_{\Vert z \Vert \to\infty}{\psi(z)}=0$, and 2) the density functions $\mathcal{Q}_t(z)$ is bounded, 
When the ansatz $\psi(z)$ has compact support, $\psi\in\mathscr{C}_c^{\infty}(\mathbb{R}^{\mathrm{D}_{\rm{LV}}})$, and~\Cref{assump:boundedCond} hold; the optimization objective for the ansatz $\psi(z)$ is expressed as:
\begin{equation}\label{eq:objective4VelocityField1}
\begin{aligned}
    \mathop{\arg\max}_{\psi(z)}~~\mathbb{E}_{\mathcal{Q}_t(z)}\left[ \psi(z)^\top\nabla_z\log{\mathcal{P}(z|x)}+\nabla_z\cdot \psi(z) \right].%\\
    %&\quad-\frac{1}{2}\mathbb{E}_{\mathcal{Q}_t(z)}\left[\psi^\top_\varphi(z)\psi_\varphi(z)\right].
\end{aligned}
\end{equation}
\end{theorem}

Notably, the learning objective defined in~\Cref{eq:objective4VelocityField1} is inherently ill-posed due to the infinite possible forms of $\psi(z)$. To address this, we propose the following theorem, which provides a closed-form expression by constraining $\psi(z)$ within the RKHS:
\begin{theorem}
Let the ansatz $\psi(z)$ be confined to the $\mathrm{D}_{\rm{LV}}$-dimensional RKHS $\mathcal{H}^{\mathrm{D}_{\rm{LV}}}$, i.e., $\psi(z) \in \mathcal{H}^{\mathrm{D}_{\rm{LV}}}$, where the corresponding kernel function $K: \mathbb{R}^{\mathrm{D}_{\rm{LV}}} \to \mathbb{R}^{\mathrm{D}_{\rm{LV}}}$ satisfies the boundary condition $\lim_{\| z \| \to \infty} K(z', z) = 0$. Under these conditions, the ansatz $\psi_{\text{RKHS}}(z)$ within RKHS can be expressed as follows:
% When the ansatz $\psi(z)$ is restricted within the $\mathrm{D}_{\rm{LV}}$-dimensional reproducing kernel Hilbert space (RKHS) $\mathcal{H}^{\mathrm{D}_{\rm{LV}}}$, i.e. $\psi(z)\in\mathcal{H}^{\mathrm{D}_{\rm{LV}}}$, and the corresponding kernel function $K(z',z):\mathbb{R}^{\mathrm{D}_{\rm{LV}}}\to \mathbb{R}^{\mathrm{D}_{\rm{LV}}}$ satisfies the boundary condition $\lim_{\Vert z \Vert \to \infty}{K(z', z)=0}$, for example, radius basis function, the ansatz $\psi_{\text{RKHS}}(z)$ can be given as follows:
\begin{equation}\label{eq:RKHSExpression}
\begin{aligned}
     & \psi_{\text{RKHS}}(z) \\
    = & \mathbb{E}_{\mathcal{Q}_t(z')}\left[K^\top(z',z)\nabla_{z'}\log{\mathcal{P}(z'|x)}+\nabla_{z'}K(z',z) \right].
\end{aligned}
\end{equation}
\end{theorem}

On this basis,~\Cref{eq:objective4VelocityFieldApprox} can be reformulated as follows based on~\Cref{eq:RKHSExpression} with the help of forward Euler's method~\cite{butcher2016numerical}:
\begin{equation}\label{eq:objective4VelocityFieldApprox2}
\begin{aligned}
    & z_{i,t+\varepsilon}= z_{i,t} 
    + \varepsilon \left\{ \nabla_z\log{\mathcal{P}(z|x)} \right.\\
 &  \left.+\mathbb{E}_{\mathcal{Q}_t(z')}\left[K^\top(z',z)\nabla_{z'}\log{\mathcal{P}(z'|x)}+\nabla_{z'}K(z',z) \right]\right\}\vert_{z=z_{i,t}},
% \dfrac{\mathrm{d}z}{\mathrm{d}t} = \psi_\varphi(z),
\end{aligned}
\end{equation}
where $\mathbb{E}_{\mathcal{Q}_t(z)}$ can be approximated by $\mathbb{E}_{\mathscr{Q}_t(z)}$. Besides, for simplicity, in this paper, we use the radius basis function (RBF) kernel function as $ K(z,z') $ to promise assumption $\psi\in\mathscr{C}_c^{\infty}(\mathbb{R}^{\mathrm{D}_{\rm{LV}}})$ holds:
\begin{equation}
    K(z,z')  \coloneqq \exp(-\frac{\Vert z -z' \Vert_2^2}{2 h}),
\end{equation}
where $h$ is the bandwidth, which is set as the median value of $\{ z_i \}_{i=1}^{\mathrm{M}}$ unless stated otherwise~\cite{gretton2012kernel,liu2016stein}. Additionally, the values of $z'$ and $z$ are identical, with the prime notation on $z'$ serving solely to distinguish the variable with respect to which the derivative is taken. Notably, in this procedure, the inference of the posterior distribution $\mathcal{P}(z|x)$ for the latent variable $z$ is achieved by simulating an \underline{Inf}inite-horizon \underline{O}ptimal control problem. Accordingly, we name our novel $\mathcal{P}(z|x)$ inference algorithm the InfO algorithm.

\begin{algorithm}[htbp]% \label{algo:odeSimulationOptimalControl}
\caption{InfO Algorithm for $\mathcal{P}(z|x)$ Inference.}\label{algo:odeSimulationOptimalControl}
\begin{algorithmic}[1]
\State \textbf{Input:} 
Target density function: $\mathcal{P}(z|x)$, intial sample for $\mathcal{Q}_0(z)=\frac{1}{\mathrm{M}}\sum_{i=1}^{\mathrm{M}}{\delta(z-z_i)}$ at time $t=0$: $\{ z_i \}_{i=1}^{\mathrm{M}}$, end time: $\mathrm{T}$, step size for $\phi(z)$ perturbation simulation: $\varepsilon$. % , learning rate for $\psi_\varphi(z)$ optimization: $\eta$, and iteration time for $\psi_\varphi(z)$ optimization: $\widehat{\mathrm{T}}$.
% Train Dataset: $\{x_{b}, y_{b}\}\vert_{b=1}^{\mathrm{N}_{\text{train}}}{\in\mathcal{D}_{\text{train}}}$, Validate Dataset: $\{x_{b}, y_{b}\}\vert_{b=1}^{\mathrm{N}_{\text{valid}}}{\in\mathcal{D}_{\text{valid}}}$, Test dataset: $\{x_{b}, y_{b}\}\vert_{b=1}^{\mathrm{N}_{\text{test}}}{\in\mathcal{D}_{\text{test}}}$.

% \State \textbf{Parameter:} parameter for $\psi_\varphi(z)$: $\varphi$.
% \State $\varphi_0 \leftarrow \varphi$
\For{$t = 0$ \textbf{to} $\mathrm{T}-1$}{\Comment{Optimal Control Simulation in Weak Sense}}
% \State $\widehat{\varphi}_0 \leftarrow \varphi_t$\;

% \For{$\tau=1$ \textbf{to} $\widehat{\mathrm{T}}-1$ }
% \State $\mathcal{L}_{\text{IPS}}(\widehat{\varphi}_\tau),\mathcal{L}_{\text{IBC}}(\widehat{\varphi}_\tau) \leftarrow \text{Eq.}~\eqref{eq:defOfIPSandIBC}$\;
% \State Obtain the gradient: $\nabla_{\widehat{\varphi}_\tau}\mathcal{L}_{\text{IPS}}(\widehat{\varphi}_\tau)$, $\nabla_{\widehat{\varphi}_\tau}\mathcal{L}_{\text{IBC}}(\widehat{\varphi}_\tau)$ \;
% \State $\alpha \leftarrow \text{Eq.}~$\Cref{eq:39a,eq:39b,eq:39c}\;
% \State $\begin{aligned} \nabla_{\widehat{\varphi}_\tau}\mathcal{L}_{\text{REC}} (\widehat{\varphi}_\tau)\leftarrow &\alpha\nabla_{\widehat{\varphi}_\tau}\mathcal{L}_{\text{IPS}}(\widehat{\varphi}_\tau) \\ &~~+ (1-\alpha)\nabla_{\widehat{\varphi}_\tau}\mathcal{L}_{\text{IBC}}(\widehat{\varphi}_\tau)\end{aligned}$\;
% \State $\widehat{\varphi}_{\tau+1}\leftarrow \text{Eq.}~\eqref{eq:varPhiUpdateForAnsatz}$\;
% \EndFor
% \State $\varphi_t \leftarrow \widehat{\varphi}_{\widehat{\mathrm{T}}}$\;
% \State $q_{\varphi_t}(z)\leftarrow \text{Eq.}~\eqref{eq:objForVarPhiTraining} $\;
% \State $z_{i,t+1}\leftarrow z_{i,t}+\varepsilon \psi_{\varphi_t}(z_{i,t})$ for $i=1,\dots,\mathrm{M}$\;
\State $z_{i}\leftarrow \text{Eq.}~\eqref{eq:objective4VelocityFieldApprox2}$
\EndFor

\State \textbf{Output:} $\{z_{i,\mathrm{T}}\}_{i=1}^{\mathrm{M}}$\;

\end{algorithmic}
\end{algorithm}
% \begin{equation}
    
% \end{equation}
\subsection{InfO-EM Algorithm: A Novel EM Algorithm for PLVMs}% \label{subsec:}

% \newpage

% Finally, we can summarize the corresponding algorithm to infer latent variable $z$ as~\Cref{algo:odeSimulationOptimalControl} shows.

% \subsection{InfO-EM Algorithm for PLVM Training}\label{subsec:PLVMTrainingAlgorithm}
\subsubsection{Procedure of InfO-EM Algorithm}
While the previous subsection effectively addresses the latent variable inference issue, it does not explicitly address the training process of the PLVM. To bridge this gap, we first present the update equation for optimizing $\theta$ based on the inferred $\mathcal{Q}_{\mathrm{T}}(z)$ at time $\mathrm{T}$, as outlined in~\Cref{algo:odeSimulationOptimalControl}. 
Accordingly, since we approximate probability distribution $\mathcal{Q}_{\mathrm{T}}(z)$ via empirical measure $\mathscr{Q}_{\mathrm{T}}(z)$, the learning objective for the M-step in~\Cref{eq:EMFunctionIteration} can be reformulated as follows:
\begin{equation}
\begin{aligned}
    \theta & =\mathop{\arg\min}_{\theta}\indent \mathbb{E}_{\mathcal{Q}_{\mathrm{T}}(z)}[\log{p_\theta(x|z)}]
    \\
    % & \approx \mathop{\arg\min}_{\theta}\indent \mathbb{E}_{\mathscr{Q}_{\mathrm{T}}(z)}[\log{p_\theta(x|z)}]\\
   &\overset{\text{(i)}}{\approx}\mathop{\arg\min}_{\theta}\indent\dfrac{1}{\mathrm{M}}\sum_{i=1}^{\mathrm{M}}{\log{p_\theta(x_i|z_{i,\mathrm{T}})}},
   \end{aligned}
\end{equation}
where `(i)' is based on the selectivity of the Dirac delta measure. 
\iffalse
% Accordingly, the parameter optimization procedure for the M-step, derived from the learning objective in~\Cref{eq:EMFunctionIteration}, can be expressed as follows, with learning rate $\xi$:  
\begin{equation}
    \theta_{\tau+1} = \theta_{\tau} + \xi \times \nabla_\theta \mathbb{E}_{\mathcal{Q}(z)}[\log{p_{\theta}(x|z)}]|_{\mathcal{Q}(z)=\mathcal{Q}_{\mathrm{T}}(z)} ,
\end{equation}
which can be further reformulated, as $x$ is supported on $\mathbb{R}^{\mathrm{D}_{\rm{obs}}}$, as follows:
\begin{equation}\label{eq:gradDescentMStep}
    \theta_{\tau+1} = \theta_{\tau} + \dfrac{\xi}{\mathrm{M}} \times \sum_{i=1}^{\mathrm{M}}{\left[ -\nabla_\theta\Vert x - \widehat{x}_i \Vert_2^2 \right]},
\end{equation}
and $\widehat{x}_i$ is the predicted value of $x$ with the help of $p_\theta(x|z)$. 
\fi
Building on this foundation, we summarize our proposed algorithm for PLVM learning in~\Cref{algo:IntoEMAlgorithmForPLVM}. Following~\Cref{subsubsec:OCProblemImplementation}, we name the proposed EM algorithm the \underline{Inf}inite-horizon \underline{O}ptimal-Control-based EM algorithm, abbreviated as the `InfO-EM' algorithm. The PLVM trained using the InfO-EM algorithm is referred to as the `InfO-PLVM'.
% Consequently, we then propose our algorithm named \underline{In}finite-\underline{T}ime-\underline{O}ptimal-Control-based EM algorithm, which is abbreviated to `InfO-EM' algorithm, for PLVM training.
\begin{algorithm}% \label{algo:odeSimulationOptimalControl}
\caption{InfO-EM Algorithm for PLVM Training.}\label{algo:IntoEMAlgorithmForPLVM}
\begin{algorithmic}[1]
\State \textbf{Input:}
Prior distribution: $\mathcal{P}(z)$, PLVM with parameter $\theta$: $p_\theta(x|z)$, intial sample for $\mathcal{Q}_0(z)=\frac{1}{\mathrm{M}}\sum_{i=1}^{\mathrm{M}}{\delta(z-z_{i,0})}$ at time $t=0$: $\{ z_{i,0} \}_{i=1}^{\mathrm{M}}$, end time: $\mathrm{T}$, and step size: $\varepsilon$. 
% learning rate for $\psi_\varphi(z)$ optimization: $\eta$, and iteration time for $\psi_\varphi(z)$ optimization: $\widehat{\mathrm{T}}$.

% Prior distribution of latent variable $z$: $\mathcal{P}(z)$, PLVM with parameter $\theta$: $p_\theta(x|z)$, intial sample $\{ z_i \}_{i=1}^{\mathrm{M}}$ for $\mathcal{Q}_0(z)=\sum_{i=1}^{\mathrm{M}}{\delta(z-z_i)}$, simulation time $\mathrm{T}$, and iterative time $\mathcal{E}$. 
% Target density function $\mathcal{P}(z|x)$, intial sample $\{ z_i \}_{i=1}^{\mathrm{M}}$ for $\mathcal{Q}_0(z)=\sum_{i=1}^{\mathrm{M}}{\delta(z-z_i)}$ at time $t=0$, end time $\mathrm{T}$, step size for $\{ z_i \}_{i=1}^{\mathrm{M}}$ evolution $\varepsilon$.
% Train Dataset: $\{x_{b}, y_{b}\}\vert_{b=1}^{\mathrm{N}_{\text{train}}}{\in\mathcal{D}_{\text{train}}}$, Validate Dataset: $\{x_{b}, y_{b}\}\vert_{b=1}^{\mathrm{N}_{\text{valid}}}{\in\mathcal{D}_{\text{valid}}}$, Test dataset: $\{x_{b}, y_{b}\}\vert_{b=1}^{\mathrm{N}_{\text{test}}}{\in\mathcal{D}_{\text{test}}}$.

\State \textbf{Parameter:} $\theta$.

\For{$e=1$ \textbf{to} $\mathcal{E}$ }% {\Comment{}}
\State $\{z_{i,\mathrm{T}}^{e}\}_{i=1}^{\mathrm{M}}\leftarrow\text{Algorithm~}\ref{algo:odeSimulationOptimalControl}$\;{\Comment{E-Step}}
\State $\theta^{e}\leftarrow\text{Eq.~}\eqref{eq:objective4VelocityFieldApprox2}$\;{\Comment{M-Step}}
% \State \;
\EndFor

\State \textbf{Output:} An optimized $p_{\theta}(x|z)\vert_{\theta=\theta^{\mathcal{E}}}$.

\end{algorithmic}
\end{algorithm}

\subsubsection{Convergence Analysis}

% Before proving the convergence, we first introduce the following two lemmas: % function kernel, 

% Based on these lemmas, we can prove the convergence of the proposed InfO-EM algorithm. Notably, similar to the spirits from other EM algorithms, in this paper, we consider the definition of convergence as follows: A sequence $\{a_1, a_2, ..., a_K\}$ is said to be convergent if there exists a real number $\mathcal{G}$ such that for any given positive number $\gamma$ ($\gamma > 0$), there exists a positive integer $N$, such that for all indices $n$ greater than $N$, the corresponding terms $a_n, n\ge N$ satisfy the inequality $\vert a_n - \mathcal{G} \vert < \gamma$. On this basis, we can check whether the sequence is monotonically increasing/decreasing and upper/lower bounded by a constant based on the celebrated monotonic convergence theorem. Consequently, we can propose the following theorem regarding the InfO-EM algorithm:
% Based on the aforementioned lemmas, we now prove the convergence of the proposed InfO-EM algorithm. 
Consistent with the convergence definitions used in traditional EM algorithms (see Section 9.2, Theorem 9.2 of reference \cite{li2023machine}), we adopt the following definition of convergence:  A sequence $\{a_1, a_2, \dots, a_{\mathcal{E}}\}$ is said to converge if there exists a real number $\mathcal{R}$ such that for any given $\gamma > 0$, there exists a positive integer $N$ such that for all $n \geq N$, the terms $a_n$ satisfy the inequality $|a_n - \mathcal{R}| < \gamma$.  To verify the convergence of a sequence, we examine whether the sequence is monotonically increasing or decreasing and whether it is upper or lower bounded by a constant. This verification is performed using the celebrated monotone convergence theorem (see Theorem 2.14 of reference~\cite{folland1999real}). With this framework, we first present the following theorem for the convergence of the InfO-EM algorithm:
\begin{theorem}\label{thm:ConvergenceAlgorithmThm}
    The InfO-EM algorithm, as summarized in~\Cref{algo:IntoEMAlgorithmForPLVM}, is guaranteed to converge provided that the step size $\varepsilon$ and learning rate $\xi$ are sufficiently small.  
    \end{theorem}

\section{Experimental Results}\label{sec:exPerimentalResults}
\begin{figure*}[htbp]
   % \vspace{-0.5cm}
    \centering
    % figures\cut_sampled_results.pdf
    \subfigure[Trajectory of $\mathcal{Q}_t(z)$ along time $t$, where $\mathcal{P}(z|x)\propto \mathcal{N}(-3, 0.5^2)$.]{\includegraphics[width=0.325\linewidth]{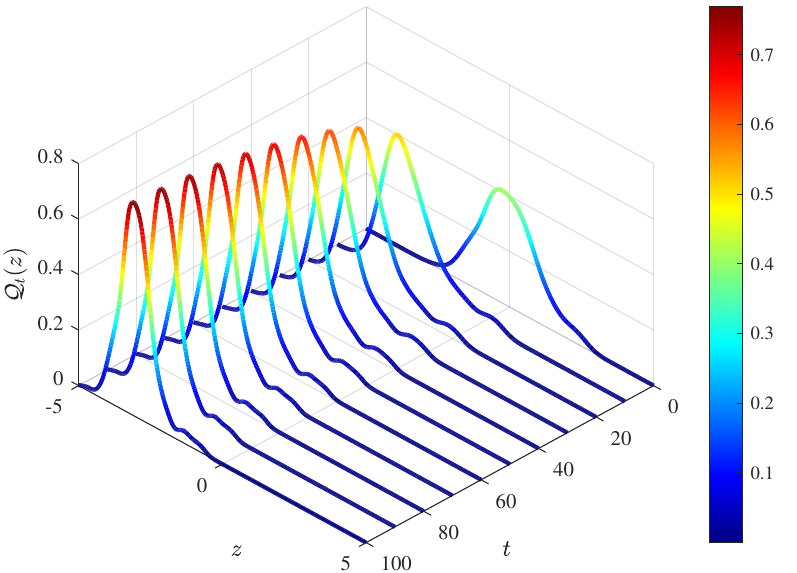}}
    \subfigure[Trajectory of $\mathcal{Q}_t(z)$  along time $t$, where $\mathcal{P}(z|x)\propto St(9,1.5,0.5)$.]{\includegraphics[width=0.325\linewidth]{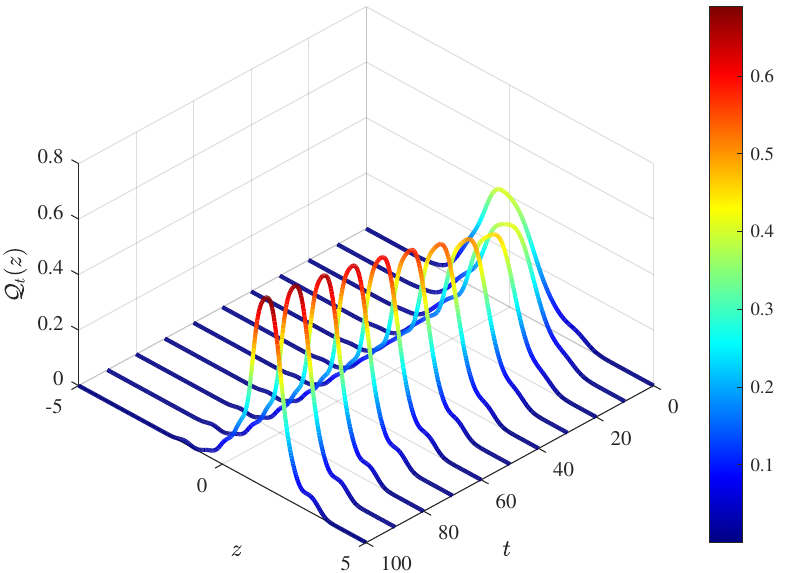}}
    \subfigure[Trajectory of $\mathcal{Q}_t(z)$  along time $t$, where $\mathcal{P}(z|x)\propto \frac{1}{2} \mathcal{N}(-2, 0.5^2)+\frac{1}{2} \mathcal{N}(2, 0.5^2)$.]{\includegraphics[width=0.325\linewidth]{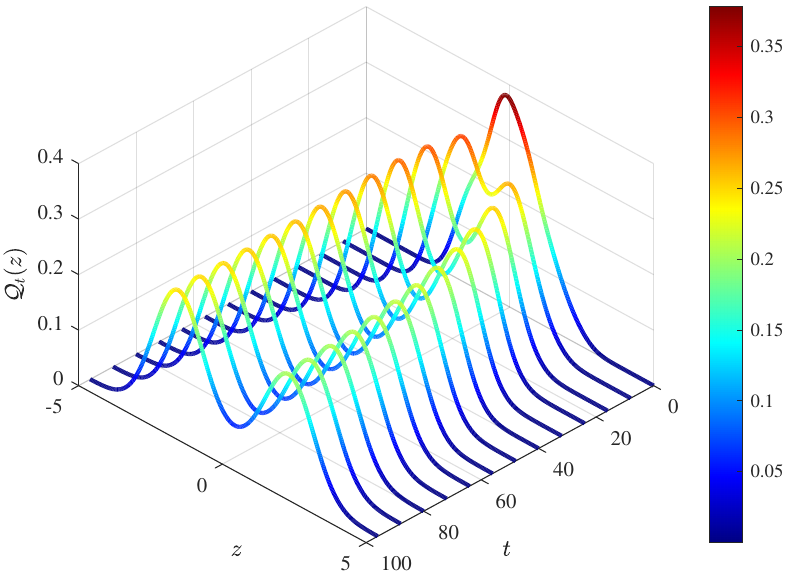}}
        \subfigure[The $\mathbb{S}\left(\mathcal{Q}_t(z),\mathcal{P}(z|x)\right)$ along time $t$, where $\mathcal{P}(z|x)\propto \mathcal{N}(-3, 0.5^2)$.]{\includegraphics[width=0.325\linewidth]{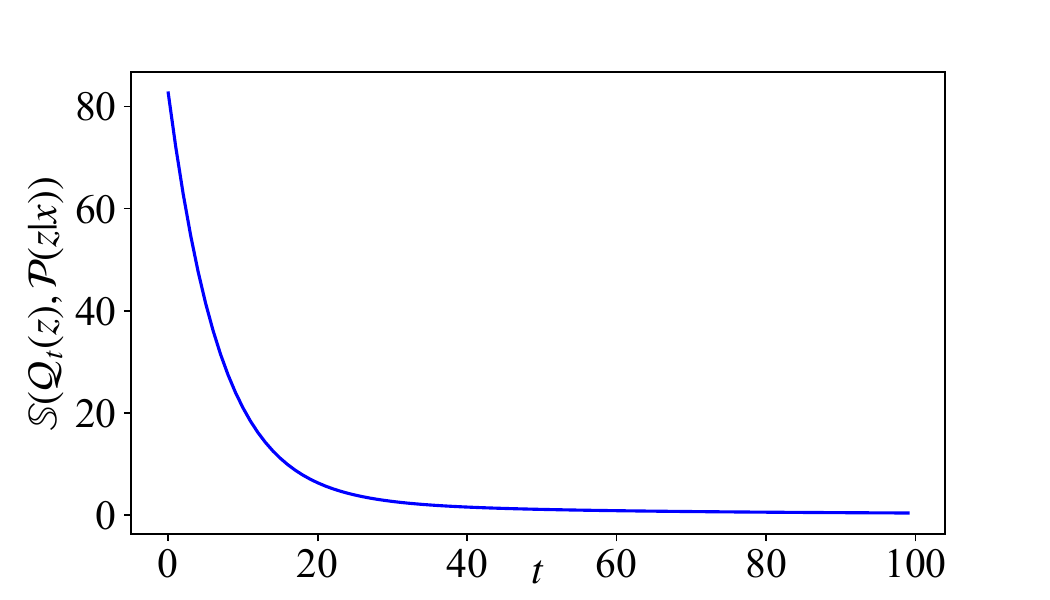}}
    \subfigure[The $\mathbb{S}\left(\mathcal{Q}_t(z),\mathcal{P}(z|x)\right)$ along time $t$, where $\mathcal{P}(z|x)\propto St(9,1.5,0.5)$.]{\includegraphics[width=0.325\linewidth]{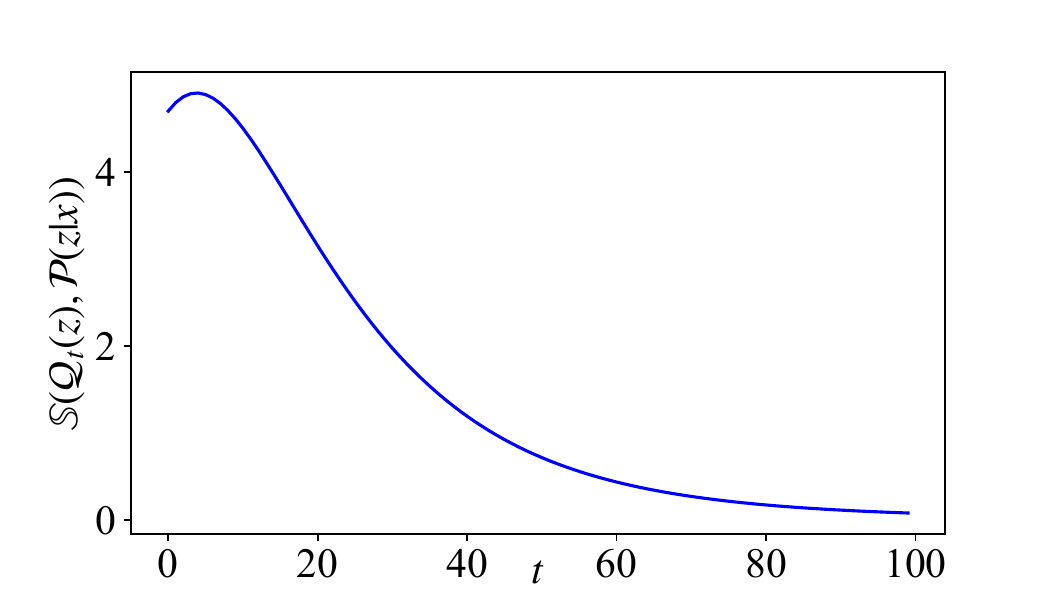}}
    \subfigure[The $\mathbb{S}\left(\mathcal{Q}_t(z),\mathcal{P}(z|x)\right)$ along time $t$, where $\mathcal{P}(z|x)\propto \frac{1}{2} \mathcal{N}(-2, 0.5^2)+\frac{1}{2} \mathcal{N}(2, 0.5^2)$.]{\includegraphics[width=0.325\linewidth]{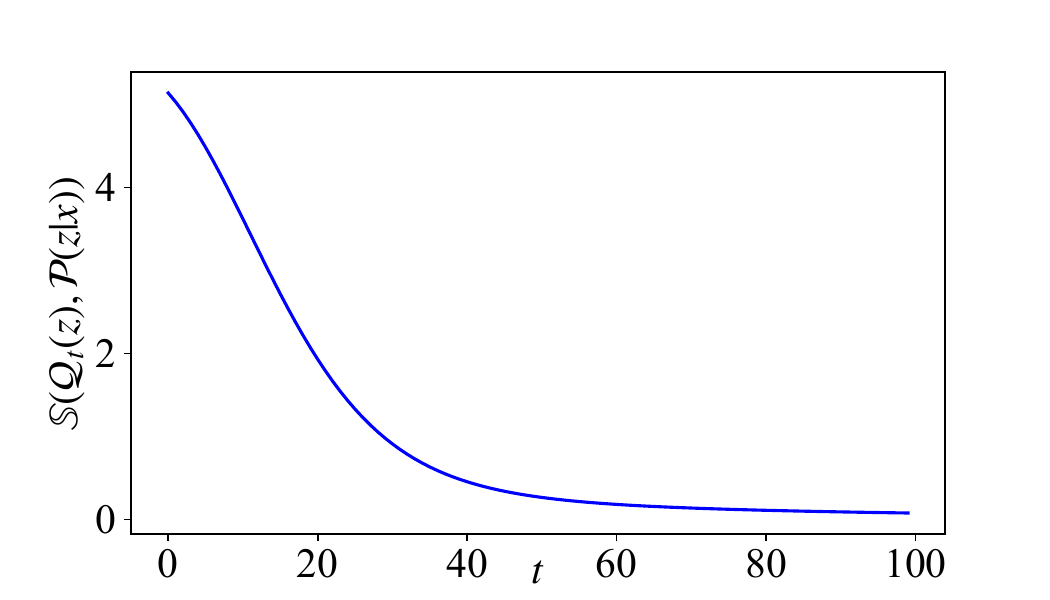}}
   
    \caption{The evolution trajectory of $\mathcal{Q}_t(z)$ (estimated by kernel density estimation with Gaussian kernel, whose bandwidth is selected by Scott's method), and the corresponding KSD value $\mathbb{S}(\mathcal{Q}_t(z), \mathcal{P}(z|x))$ along time $t$.% Comparison results between $\mathcal{P}(z|x)/p_{\theta,\text{NF}}(z|x)$ with $\mathcal{Q}(z)$ along iterative time $\tau$. %  $q(\vec{a})$.
    %. (b) Impact of entropic regularization strength ε. (c) Impact of PFOR strength γ (×103). (d) Impact of RMPR strength κ.
    }\label{fig:sensInfLR}
   % \vspace{-0.1cm}
\end{figure*}
% \vspace{-0.3cm}
In this section, we conduct experiments to answer the following four main research questions (RQs) empirically:
\\\textbf{RQ1 (Effectiveness):} \textit{Does the InfO algorithm take effect?}
\\\textbf{RQ2 (Accuracy):} \textit{What's the posterior distribution inference accuracy of the InfO algorithm compared with other variational inference approaches?}
\\\textbf{RQ3 (Performance):} \textit{What's the performance of PLVM trained by InfO-EM algorithm compared with other PLVMs?}
\\\textbf{RQ4 (Convergence):} \textit{Does the InfO-EM algorithm converge?}
In addition, we further study the following RQ in our supplementary material due to the page limit of the main content:
% \\\textbf{RQ5 (Sensitivity):} \textit{How does the performance of InfO-PLVM change with varying hyperparameters?}

\subsection{Posterior Distribution Approximation Trajectory Study}\label{subsec:densityEvolutionStudy}
% picture\cut_evolution_pdf\cut_gmm_evolution.pdf

In this subsection, we address $\textbf{RQ1}$: ``Does the InfO algorithm take effect?'' To answer this, we conduct a qualitative experiment visualizing the density evolution trajectory. To facilitate observation of the evolution process, we consider three types of one-dimensional distributions: the Gaussian distribution ($\mathcal{N}(\mu, \sigma^2)$, where $\mu$ and $\sigma$ denote the location and scale, respectively), the Student's-$t$ distribution ($St(\nu, \mu, \sigma)$, where $\nu$, $\mu$, and $\sigma$ represent the degree of freedom, location, and scale, respectively), and the Mixture of Gaussian distribution. In addition, the initial distribution $\mathcal{Q}_0(z)$ is set to a standard Gaussian distribution $\mathcal{N}(0, 1)$ for fairness,.  

For evaluation, we employ the widely used kernelized Stein discrepancy (KSD) $\mathbb{S}$, defined in~\Cref{eq:steinDefinition,eq:UComputational}, as the evaluation metric. According to references~\cite{liu2016kernelized}, a smaller KSD value indicates a higher approximation accuracy, making it an appropriate measure for this comparison. In addition, to promise the nonnegative property of $\mathbb{S}\left(\mathcal{Q}(z),\mathcal{P}(z|x)\right)$, we design the `V-statistics' defined in~\Cref{eq:UComputational} for $\mathbb{S}\left(\mathcal{Q}(z),\mathcal{P}(z|x)\right)$ computation as shown in the last line of~\Cref{eq:steinDefinition}.
\iffalse
In this subsection, we further investigate the efficacy of the proposed latent variable inference strategy and answer $\textbf{RQ2}$ \textit{What's the accuracy of the InfO algorithm compared with other approaches?}. To this end, we consider three types of special distributions namely `Mixture of Gaussian' (MoG), `mixture of ring' (MoR), and `two moon' (TM) as posterior distribution $\mathcal{P}(z|x)$. The corresponding PDF of these distributions are given in~\Cref{fig:contourResult} (a) to (c).  Based on this, we consider comparing the inferred result of InfO algorithm with the scenarios that specify $\mathcal{Q}(z)$ by the unimodal Gaussian distribution (Gauss) and Gaussian mixture model (GMM). We choose the celebrated kernelized Stein discrepancy (KSD) $\mathbb{S}$ defined by~\Cref{eq:steinDefinition,eq:UComputational} as our evaluation metric based on~\cite{liu2016kernelized}, and a smaller KSD indicates a higher approximation accuracy. 
\fi
\begin{subequations}
    \begin{align}
       & \begin{aligned}
       & \mathbb{S}\left(\mathcal{Q}(z),\mathcal{P}(z|x)\right)  \coloneqq  \mathbb{E}_{z,z'\sim \mathcal{Q}(z)}\left[\mathcal{V}_{\mathcal{P}(z|x)}(z,z')\right]\\
        & \approx \mathbb{E}_{z,z'\sim \mathscr{Q}(z)}\left[\mathcal{V}_{\mathcal{P}(z|x)}(z,z')\right] \approx \dfrac{1}{\mathrm{M}^2}\sum_{i=1}^{\mathrm{M}}{\sum_{j=1}^{\mathrm{M}}{  \mathcal{V}_{\mathcal{P}(z|x)}\left(z_i,z_j\right)}} ,
        \end{aligned} \label{eq:steinDefinition} \\
       & \begin{aligned}
     &   \mathcal{V}_{\mathcal{P}(z|x)}(z,z') \\
     &  \coloneqq \left[\nabla_z\log{\mathcal{P}(z|x)}\right]^\top K(z,z')\left[\nabla_{z'}\log{\mathcal{P}(z'|x)}\right]\\
       & + \left[\nabla_z\log{\mathcal{P}(z|x)}\right]^\top \nabla_{z'}K(z,z') +  \mathrm{Trace}\left(\nabla_{z,z'}K(z,z')\right) \\
       & +\left[\nabla_{z}K(z,z')\right]^\top[\nabla_{z'}\log{\mathcal{P}(z'|x)}].
           \end{aligned} \label{eq:UComputational}
    \end{align}
    \end{subequations}
The evolution trajectories of $\mathcal{Q}_t(z)$ over time $t$ are visualized in~\Cref{fig:sensInfLR} (a) to (c), and the corresponding KSD values $\mathbb{S}(\mathcal{Q}_t(z),\mathcal{P}(z|x))$ are shown in~\Cref{fig:sensInfLR} (d) to (f). The following observations can be made from these figures:
\begin{enumerate}[leftmargin=*] %[leftmargin=*]
\item In Fig.~\ref{fig:sensInfLR}(a), starting from the standard Gaussian distribution $\mathcal{Q}_0(z)$, the perturbation process progressively adjusts both the mean and variance of $\mathcal{Q}_t(z)$, illustrating the algorithm's ability to flexibly adapt the location and scale of the distribution.
\item In Fig.~\ref{fig:sensInfLR}(b), the algorithm successfully transforms $\mathcal{Q}_t(z)$ from a light-tailed to a heavy-tailed distribution, indicating its capacity to modulate tail behavior.
\item In Fig.~\ref{fig:sensInfLR}(c), $\mathcal{Q}_t(z)$ evolves from a unimodal to a bimodal distribution, demonstrating the method's capability to alter the fundamental modality of the distribution.
\item Figs.~\ref{fig:sensInfLR}(d)–(f) show that the corresponding KSD values $\mathbb{S}(\mathcal{Q}_t(z),\mathcal{P}(z|x))$ decrease steadily during the evolution process, reflecting the effectiveness of the proposed InfO algorithm in approximating the target posterior.
\end{enumerate}
In summary, these qualitative results clearly demonstrate the effectiveness of the proposed posterior inference strategy, highlighting its flexibility in adapting both the shape and characteristics of the underlying distributions.

\subsection{Posterior Distribution Inference Accuracy Comparison}\label{subsec:goodnessStudy}
\begin{figure}[!h]
    % \vspace{-0.5cm}
\centering
     % figures\cut_sampled_results.pdf 
     % picture\cut_plot
     % D:\PycharmProjects\SteinVariational\ParVI\PrecondVIOC\exper_pfg\plot_baseline\cut_plot\cut_gmm_scatter_plot_1.pdf
\subfigure[MoG.]{\includegraphics[width=0.325\linewidth]{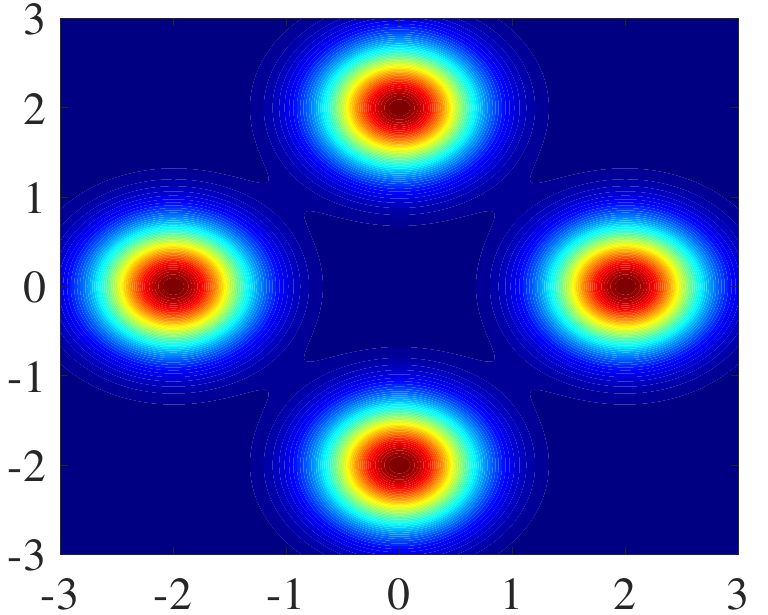}}
\subfigure[MoR.]{\includegraphics[width=0.325\linewidth]{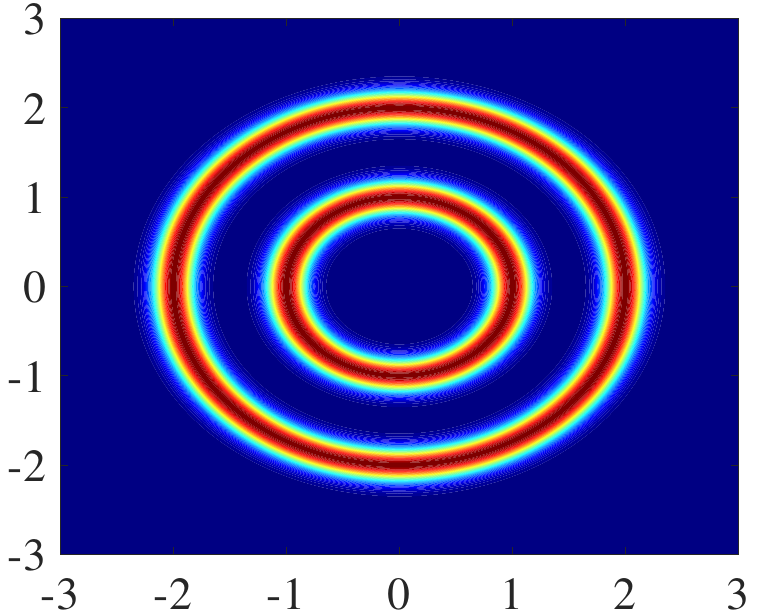}}
\subfigure[TM.]{\includegraphics[width=0.325\linewidth]{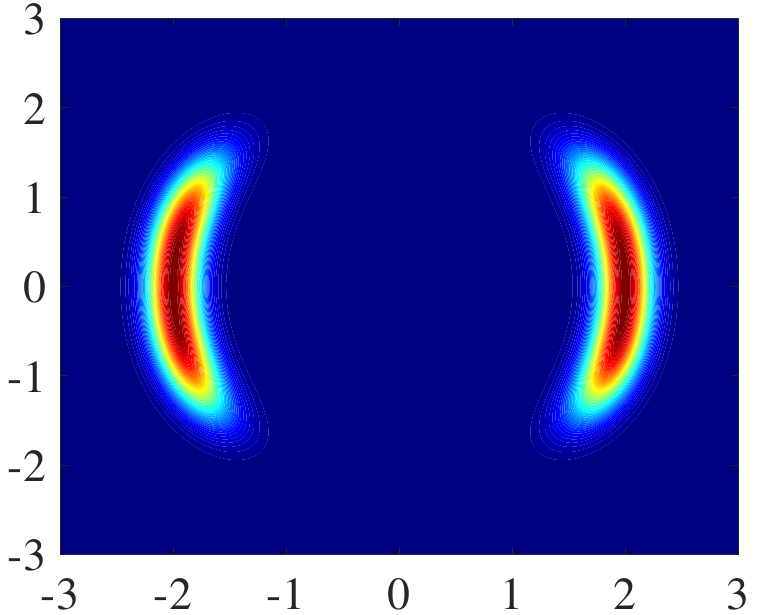}}

\subfigure[Gauss, MoG.]{\includegraphics[width=0.325\linewidth]{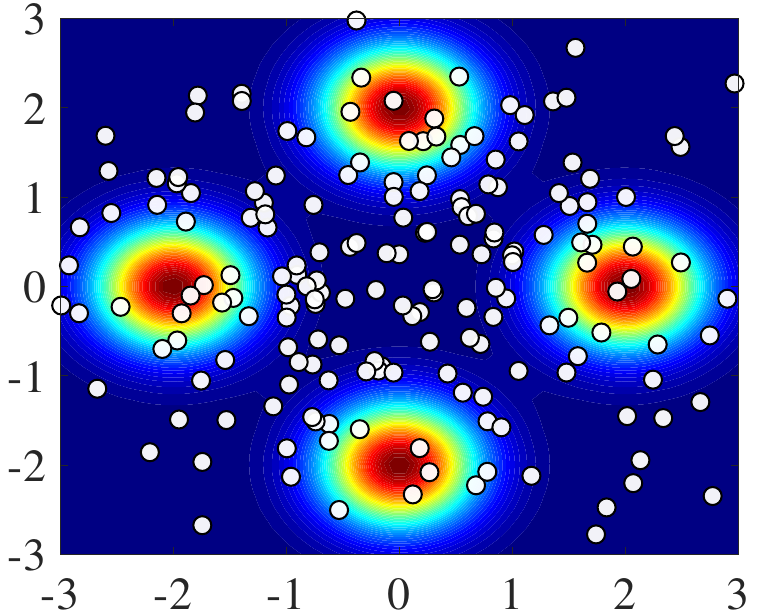}}
\subfigure[Gauss, MoR.]{\includegraphics[width=0.325\linewidth]{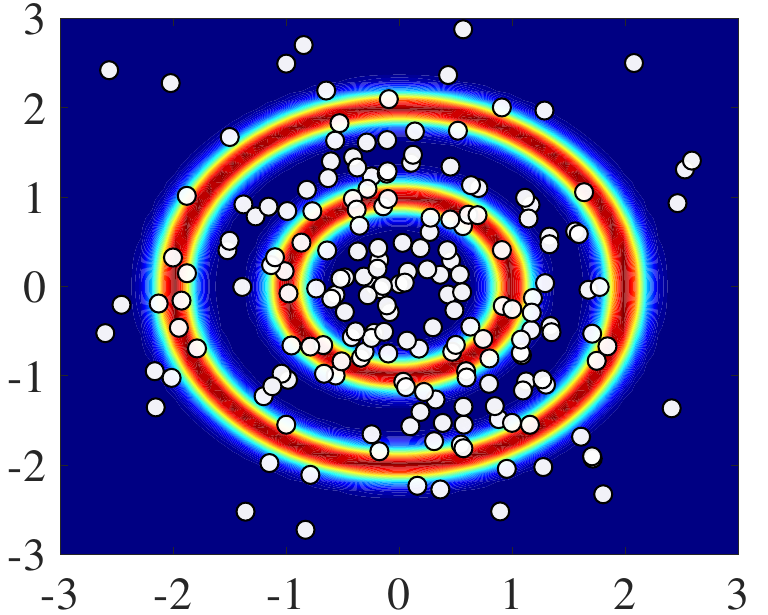}}
\subfigure[Gauss, TM.]{\includegraphics[width=0.325\linewidth]{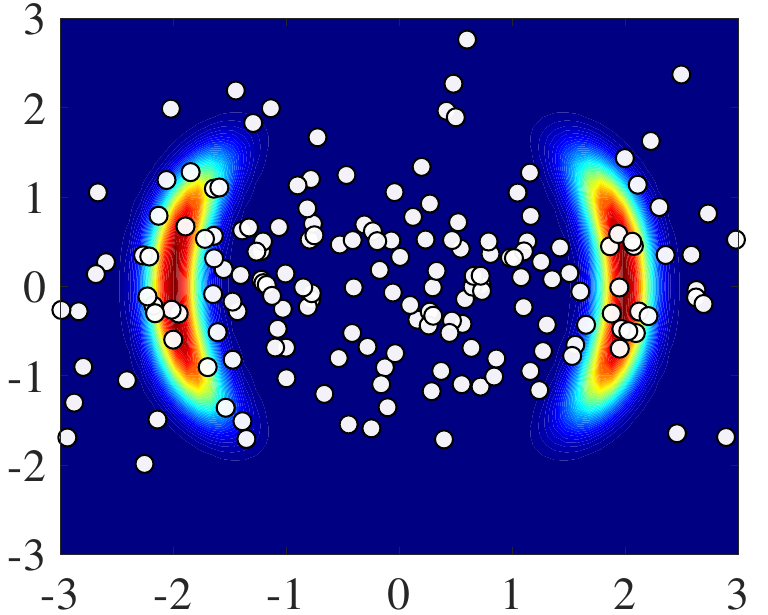}}

\subfigure[GMM, MoG.]{\includegraphics[width=0.325\linewidth]{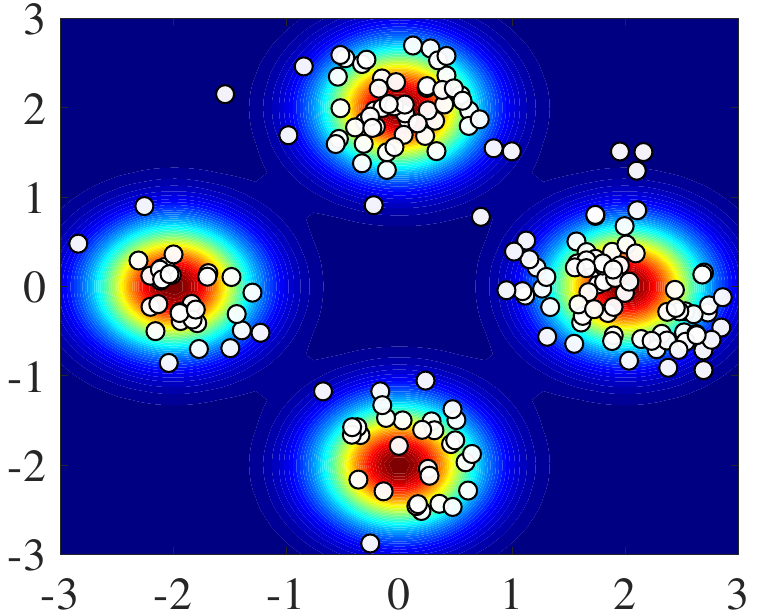}}
\subfigure[GMM, MoR.]{\includegraphics[width=0.325\linewidth]{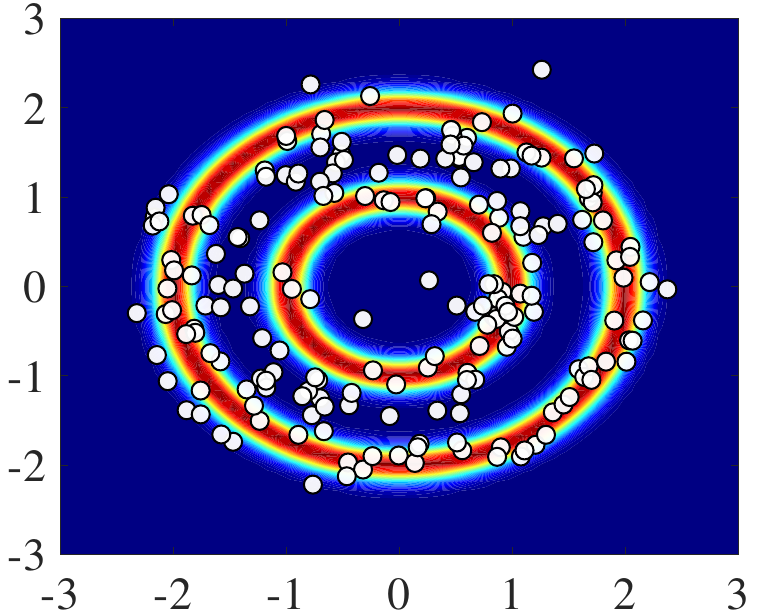}}
\subfigure[GMM, TM.]{\includegraphics[width=0.325\linewidth]{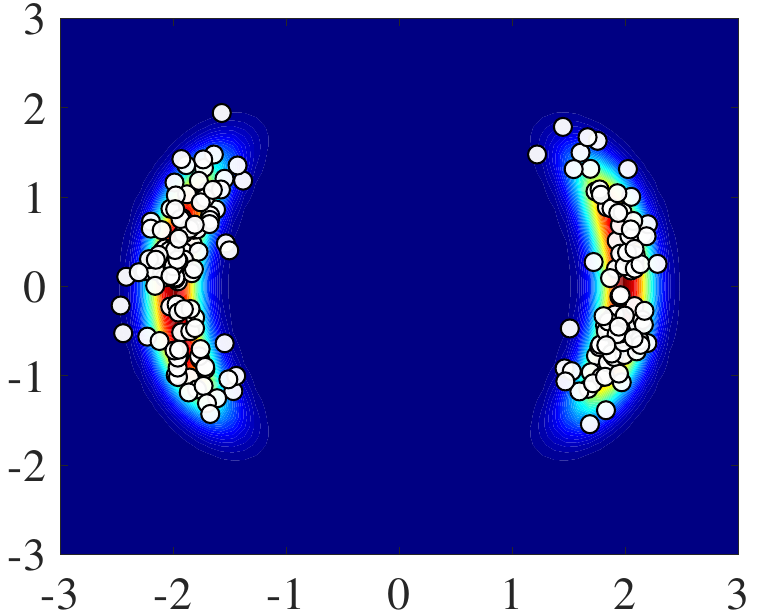}}
\subfigure[InfO, MoG, $h=0.5$.]{\includegraphics[width=0.325\linewidth]{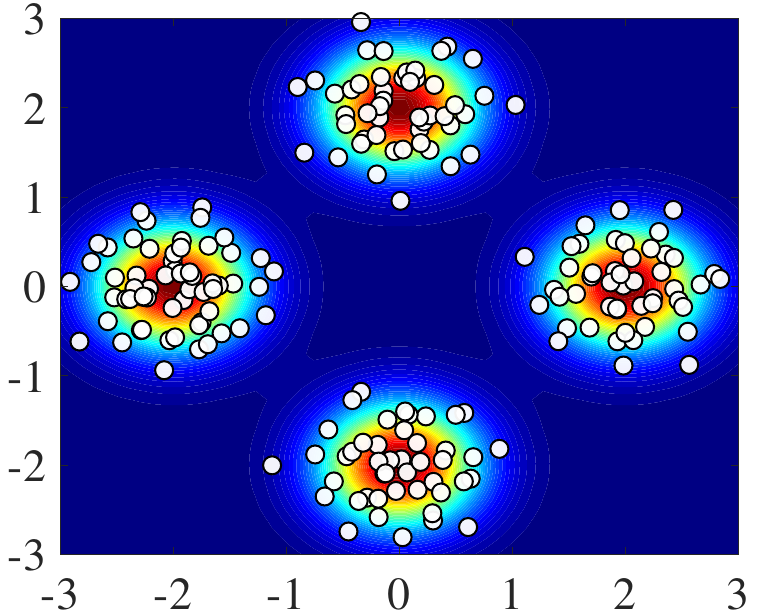}}
\subfigure[InfO, MoR, $h=1.0$.]{\includegraphics[width=0.325\linewidth]{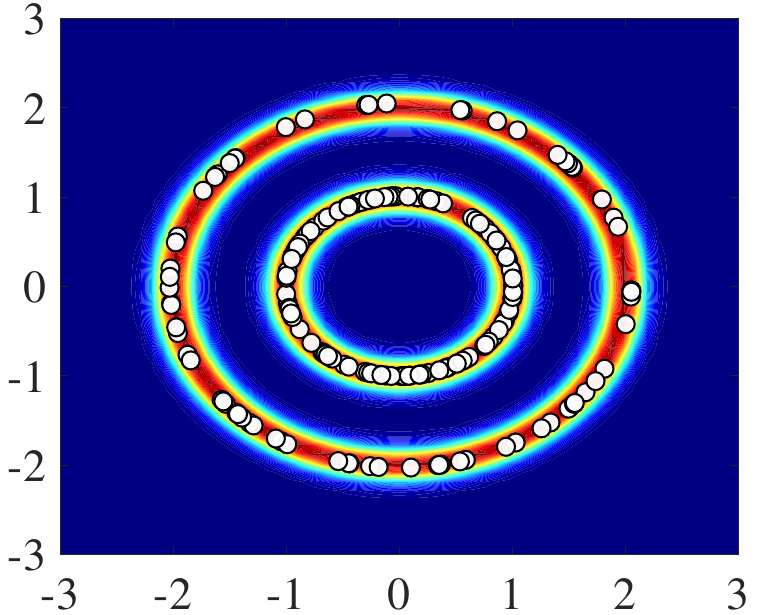}}
\subfigure[InfO, TM, $h=0.5$.]{\includegraphics[width=0.325\linewidth]{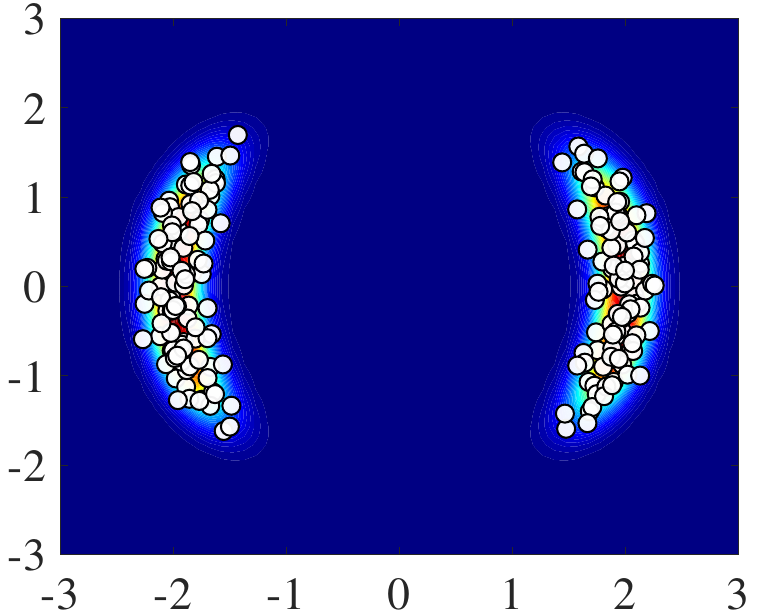}}

\caption{The contour of PDF for $\mathcal{P}(z|x)$ vary different distributions ((a) to (c)). The wihte dots are samples from $\mathcal{Q}(z)$. The inference results when $\mathcal{Q}(z)$ is specified by Gauss ((d) to (f)). The inference results when $\mathcal{Q}(z)$ is specified by an eight-element GMM ((g) to (i)). The inference results when $\mathcal{Q}(z)$ is obtained by InfO-EM algorithm ((j) to (l)). 
}\label{fig:contourResult}
\vspace{-0.5cm}
\end{figure}

In this subsection, we further evaluate the efficacy of the proposed latent variable inference strategy and address 
\textbf{RQ2}: ``What is the accuracy of the InfO algorithm compared to other approaches?'' To this end, we consider three specific types of posterior distributions, namely the Mixture of Gaussian (MoG), Mixture of Ring (MoR), and Two Moon (TM) distributions. The corresponding PDFs of these distributions are visualized in~\Cref{fig:contourResult} (a) to (c).

To assess the performance of the InfO algorithm, we compare its inferred results with scenarios where $\mathcal{Q}(z)$ is specified using an unimodal Gaussian distribution (Gauss) and a Gaussian Mixture Model (GMM). In addition to evaluating performance based on KSD, we further conduct the goodness-of-fit test using KSD as the test statistic (due to page constraints, the detailed procedure is provided in the supplementary material). During the goodness-of-fit test, the hypotheses are defined as follows, with the significance level set at $0.05$:  
\begin{itemize}[leftmargin=*] 
    \item{$H_0$: The samples $\{ z_i \}_{i=1}^{\mathrm{M}} \sim \mathcal{Q}(z)$ are drawn from $\mathcal{P}(z|x)$.}
    \item{$H_1$: The samples $\{ z_i \}_{i=1}^{\mathrm{M}} \sim \mathcal{Q}(z)$ are not drawn from $\mathcal{P}(z|x)$.}
\end{itemize}

\begin{table}[!h]
    \caption{Approximation Accuracy Comparison}\label{tab:approximationAccuracy}
    \resizebox{\columnwidth}{!}{
        \begin{threeparttable}
\begin{tabular}{l|ll|ll|ll}
\toprule
\multirow{2}{*}{Method} & \multicolumn{2}{c|}{MoG} & \multicolumn{2}{c|}{MoR} & \multicolumn{2}{c}{TM} \\ \cmidrule{2-7} 
                        &   $\mathbb{S}$           & $H_0/H_1$         &   $\mathbb{S}$            &   $H_0/H_1$       &  $\mathbb{S}$            &   $H_0/H_1$      \\ \midrule
Gauss                   & 3.92E-1$\dag$     & $H_1$    & 1.50E1$\dag$     & $H_1$    & 5.76E1$\dag$    & $H_1$   \\
GMM                     & 7.11E-2$\dag$     & $H_1$    & 2.29E0$\dag$     & $H_1$    & 4.19E-1$\dag$    & $H_1$   \\
InfO                    & \textbf{5.53E-3}     & $H_0$    & \textbf{1.99E-3}     & $H_0$    & \textbf{9.66E-3}    & $H_0$   \\ \bottomrule
\end{tabular}
\begin{tablenotes}
    \item{$\dag$ marks the variants that InfO algorithm significantly at $p$-value $<$ 0.05 over paired samples $t$-test. \textbf{Bolded} results indicate the best result.
    }
    \end{tablenotes}
    \end{threeparttable}
    }  
    \vspace{-0.5cm}
\end{table}

The results of the test are visualized in Figs.~\ref{fig:contourResult} (d) to (l) and summarized in~\Cref{tab:approximationAccuracy}. From Figs.~\ref{fig:contourResult} (d) to (f), we observe that when $\mathcal{Q}(z)$ is specified as an unimodal Gaussian (Gauss), most samples fail to fall within the high PDF regions, revealing the limitations of this model's flexibility. Subsequently, when $\mathcal{Q}(z)$ is specified as a Gaussian Mixture Model (GMM), we observe that although some samples do fall within the high PDF regions, a significant number still reside in low-density regions. This phenomenon persists even when the number of components in the GMM is increased, as shown in Figs.~\ref{fig:contourResult} (g) to (i), highlighting the continued impact of model specification constraints on approximation accuracy. Finally, when applying the proposed InfO algorithm for inferring $\mathcal{P}(z|x)$, as depicted in Figs.~\ref{fig:contourResult} (j) to (l), nearly all samples are concentrated in the high-density regions. These visual results are further supported by the quantitative findings presented in~\Cref{tab:approximationAccuracy}. Specifically, from~\Cref{tab:approximationAccuracy}, we observe that the InfO algorithm outperforms both the Gauss and GMM baselines by one to two orders of magnitude in terms of KSD and achieves superior results in the goodness-of-fit test. These observations demonstrate the efficacy of the proposed relaxation strategy, providing a clear answer to $\textbf{RQ2}$ and further validating the advantages of the InfO algorithm.
% This phenomenon indicates that our relaxation strategy of model spe   %the approximated instances may 

\begin{table*}[htbp]
    \caption{Inferential Sensor Accuracy Comparison}\label{tab:softSensorAccuracy}
   \resizebox{\linewidth}{!}{
        \begin{threeparttable}
\begin{tabular}{l|llll|llll|llll}\toprule \multirow{2}{*}{Model} & \multicolumn{4}{c|}{DC}& \multicolumn{4}{c|}{CA}& \multicolumn{4}{c}{WGS}\\ \cmidrule{2-13} & \multicolumn{1}{c}{$\text{R}^\text{2}$} & \multicolumn{1}{c}{RMSE} & \multicolumn{1}{c}{MAE} & \multicolumn{1}{c|}{MAPE} & \multicolumn{1}{c}{$\text{R}^\text{2}$} & \multicolumn{1}{c}{RMSE} & \multicolumn{1}{c}{MAE} & \multicolumn{1}{c|}{MAPE} & \multicolumn{1}{c}{$\text{R}^\text{2}$} & \multicolumn{1}{c}{RMSE} & \multicolumn{1}{c}{MAE} & \multicolumn{1}{c}{MAPE} \\ \midrule SNPLVR  & -9.19E-2$\dagger$ & 2.10E-1$\dagger$ & 1.68E-1$\dagger$ & 2.75E2$\dagger$ & -1.70E-1$\dagger$ & 7.70E-3$\dagger$ & 5.72E-3$\dagger$ & 1.95E0$\dagger$ & -3.28E-1$\dagger$ & 6.80E-1$\dagger$ & 5.46E-1$\dagger$ & 2.78E-1$\dagger$ \\ \midrule DBPSFA  & 2.50E-1$\dagger$ & 1.75E-1$\dagger$ & 1.43E-1$\dagger$ & 2.81E2$\dagger$ & -6.77E-2$\dagger$ & 7.36E-3$\dagger$ & 5.63E-3$\dagger$ & 1.93E0$\dagger$ & -3.73E4$\dagger$ & 1.15E2$\dagger$ & 1.15E2$\dagger$ & 5.85E1$\dagger$ \\ \midrule MUDVAE-SDVAE  & 2.18E-2$\dagger$ & 2.00E-1$\dagger$ & 1.60E-1$\dagger$ & 2.71E2$\dagger$ & -1.05E-3$\dagger$ & 7.13E-3$\dagger$ & 5.23E-3$\dagger$ & 1.78E0$\dagger$ & -1.60E-1$\dagger$ & 6.40E-1$\dagger$ & 5.13E-1$\dagger$ & 2.61E-1$\dagger$ \\ \midrule GMVAE  & \underline{7.96E-1}$\dagger$ & \underline{8.09E-2}$\dagger$ & \underline{6.51E-2}$\dagger$ & \underline{9.72E1}$\dagger$ & \underline{2.96E-1}$\dagger$ & \underline{6.12E-3}$\dagger$ & \underline{4.64E-3}$\dagger$ & \underline{1.59E0}$\dagger$ & \underline{7.84E-1}$\dagger$ & \underline{2.75E-1}$\dagger$ & \underline{2.20E-1}$\dagger$ & \underline{1.12E-1}$\dagger$ \\ \midrule InfO-PLVM  & \textbf{9.96E-1} & \textbf{1.26E-2} & \textbf{9.71E-3} & \textbf{1.81E1} & \textbf{7.51E-1} & \textbf{3.55E-3} & \textbf{2.78E-3} & \textbf{9.54E-1} & \textbf{9.40E-1} & \textbf{1.46E-1} & \textbf{1.17E-1} & \textbf{5.95E-2} \\ \bottomrule\end{tabular}
\begin{tablenotes}
    \item{$\dag$ marks the variants that InfO-PLVM model significantly at $p$-value $<$ 0.05 over paired samples $t$-test. \textbf{Bolded} results indicate the best in each metric. \uline{Underlined} results indicate the second best in each metric.
    }
    \end{tablenotes}
    \end{threeparttable}
}  
\vspace{-0.6cm}
\end{table*}

\subsection{Performance on Inferential Sensor Task}\label{subsec:SSPerformanceComparison}
\color{black}
% In this subsection, we aim to address \textbf{RQ3}: ``What is the performance of PLVMs trained using the InfO-EM algorithm compared to other PLVMs?'' To this end, we apply the PLVM to the inferential sensor task for industrial process to demonstrate the superiority of the InfO-PLVM. Specifically, the inferential sensor task attempt to estimate those hard-to-measure key quality variables via the easy-to-measure process variables. On this basis, we conduct the inferential sensor task on three datasets, namely the debutanizer column (DC), carbon-dioxide absorber (CA), and water-gas-shift (WGS) and compare the performance of PLVMs trained with the InfO-EM algorithm against other PLVMs in terms of prediction accuracy. More detailed information of these datasets are provided in the supplementary material.

To address \textbf{RQ3}:—``What is the performance of PLVMs trained using the InfO-EM algorithm compared to other PLVMs?''\\—we conduct a comparative study on an industrial inferential sensor task. The goal of this task is to estimate critical but hard-to-measure quality variables from easily obtainable chemical process data. We benchmark the performance on three datasets collected from real chemical processes: the debutanizer column (DC), carbon-dioxide absorber (CA), and water-gas-shift (WGS). The prediction accuracy of our InfO-EM-trained PLVM is systematically compared with that of other PLVM baselines. A detailed description of each dataset is provided in the supplementary material.
% To this end, we conduct a soft sensor experiment on the debutanizer column (DC) process, a benchmark process commonly used in the context of soft sensors, and compare the performance of PLVMs trained with the InfO-EM algorithm against other PLVMs in terms of prediction accuracy.

% To evaluate model performance on the soft sensor task, we adopt the metrics $\textrm{RMSE}$, $\textrm{R}^\text{2}$, $\textrm{MAE}$, and $\textrm{MAPE}$, as defined in~\Cref{eq:evaluationMetrics}. Here, $\mathrm{N}_{\text{test}}$ denotes the size of the testing dataset, and $\bar{x}$ represents the mean value of the target variable. For $\textrm{RMSE}$, $\textrm{MAE}$, and $\textrm{MAPE}$, smaller values indicate higher prediction accuracy, while for $\textrm{R}^2$, values closer to 1 signify better predictive performance. 
The model's predictive performance on the inferential sensor task is assessed using Root Mean Square Error (RMSE), Coefficient of Determination ($\text{R}^\text{2}$), Mean Absolute Error (MAE), and Mean Absolute Percentage Error (MAPE). The formal definitions are provided in~\Cref{eq:evaluationMetrics}, where $\mathrm{N}_{\text{test}}$ is the test set size and $\bar{x}$ is the mean of the target variable. Lower values of RMSE, MAE, and MAPE correspond to better performance, while for $\text{R}^\text{2}$, values closer to $1$ are desirable.
\begin{equation}\label{eq:evaluationMetrics}
\begin{cases}
      \textrm{RMSE} &=  \sqrt {\tfrac{1}{\mathrm{N}_{\text{test}}}\sum\limits_{l = 1}^{\mathrm{N}_{\text{test}}} {{{({x_l} - {{\hat{x}}_l})}^2}} } ,\\
   \textrm{R}^{\text{2}}   & = 1 - \tfrac{{\sum\limits_{l = 1}^{\mathrm{N}_{\text{test}}} {{{({x_l} - {{\hat{x}}_l})}^2}} }}{{\sum\limits_{l = 1}^{\mathrm{N}_{\text{test}}} {{{({x_l} - \bar x)}^2}} }}, \\
  \textrm{MAPE}   &=  \tfrac{1}{\mathrm{N}_{\text{test}}}\sum\limits_{l = 1}^{\mathrm{N}_{\text{test}}} {{{\vert \tfrac {({x_l} - {{\hat{x}}_l})}{{x_l}}\vert}}} \times 100\% \\ 
    \textrm{MAE} & = \frac{1}{\mathrm{N}_{\text{test}}}\sum\limits_{l = 1}^{\mathrm{N}_{\text{test}}} {{{\vert{x_l} - {{\hat{x}}_l}\vert}}} 
\end{cases}\end{equation}

On this basis, the following PLVMs designed for inferential sensor modeling are chosen: Deep Bayesian Probabilistic Slow Feature Analysis (DBPSFA)~\cite{9625835}
, Modified Unsupervised VAE-Supervised Deep VAE (MUDVAE-SDVAE)~\cite{xie2019supervised}, Nonlinear Probabilistic Latent Variable Regression (NPLVR)~\cite{shen2020nonlinear}, and Gaussian Mixture-Variational Autoencoder (GMVAE)~\cite{guo2021just}. Notably, the variational distribution for DBPSFA, MUDVAE-SDVAE, and NPLVR is set as the standard Gaussian distribution, and the variational distribution for GMVAE is set as the eight-modal GMM. Other detailed information about hyperparameters is provided in the supplementary material.  % The corresponding results are listed in~\Cref{tab:softSensorAccuracy}.

The results of these experiments are summarized in~\Cref{tab:softSensorAccuracy}. From~\Cref{tab:softSensorAccuracy}, we observe that the choice of variational distribution significantly affects model performance on downstream tasks, corroborating the insights discussed in~\Cref{subsec:implicitLVSpec}. Specifically, models utilizing the unimodal Gaussian distribution as the variational distribution, such as DBPSFA, MUDVAE-SDVAE, and NPLVR, exhibit suboptimal performance. In contrast, models employing a GMM for the variational distribution show notable improvements in performance, as evidenced by the results for GMVAE. 

Building on this, when we apply the InfO-EM algorithm, which relaxes the model specification, the resulting PLVM achieves the best performance across all evaluation metrics. This improvement reflects the importance of model specification relaxation for enhancing downstream task performance and further underscores the superiority of the proposed InfO-EM algorithm.

\subsection{Convergence Analysis}\label{subsec:convergenceAnalysisResult}
\begin{figure}[!h]
    \centering
    \includegraphics[width=0.35\textwidth]{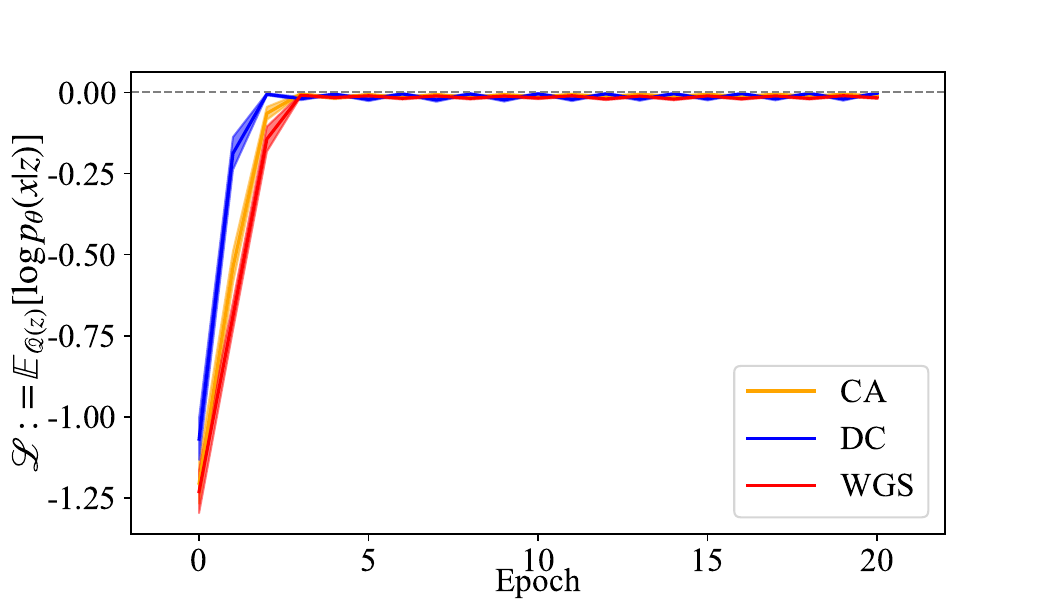}
    \caption{The convergence results of the InfO-EM algorithm on three datasets, where the shaded area indicates $\pm$0.25 times the standard deviation.}\label{fig:convergenceResult}
    \vspace{-0.5cm}
\end{figure}
In this section, we empirically validate the convergence properties of the InfO-EM algorithm, addressing \textbf{RQ4}: ``Does the InfO-EM algorithm converge?"~\Cref{fig:convergenceResult} illustrates the evolution of the expected log-likelihood, $\mathbb{E}_{\mathscr{Q}(z)}[\log p_{\theta}(x|z)]$, as a function of training epochs on the CA, DC, and WGS datasets. A clear pattern of rapid convergence is observable. In all datasets, the learning objective swiftly rises from a negative value and plateaus near the optimal value of zero in under 5 epochs. The minimal standard deviation, visualized as a tight shaded region around the mean curve, further attests to the algorithm's stable and consistent performance across runs. These empirical results, in perfect alignment with our theoretical analysis in~\Cref{thm:ConvergenceAlgorithmThm}, provide compelling evidence for the rapid and stable convergence of the InfO-EM algorithm. % These findings provide strong evidence for the reliable convergence of InfO-EM.

\section{Conclusions}\label{sec:Conclusions}
In this paper, we proposed an infinite-horizon optimal control approach to address the model specification challenge in PLVMs and enhance their performance on downstream tasks. Specifically, we represented the approximation distribution as a finite set of particles and established that their ODE-driven dynamics provide a weak solution to the continuity equation governing distributional flows. Based on this theoretical foundation, we reformulated the inference of the latent variable distribution in PLVMs into an infinite-horizon optimal control problem. This reformulation transforms the task of inferring the latent variable distribution within a predefined normalized distribution family, $\mathbb{F}$, into the problem of determining an optimal control policy within the infinite-horizon path space, $C\left([0,\infty), \mathbb{R}^{\mathrm{D}_{\rm{LV}}}\right)$. Building on this formulation, we derived the corresponding optimal control policy using Pontryagin's maximum principle and proposed a tractable ansatz to approximate this otherwise intractable control policy. Furthermore, we summarized the inference procedure, termed the InfO algorithm, introduced a novel EM algorithm for PLVM training, termed the InfO-EM algorithm, and proved the convergence properties of the InfO-EM algorithm. Finally, we conducted extensive experiments to validate the efficacy and robustness of the proposed InfO and InfO-EM algorithms.

\noindent \textbf{Limitations \& Future Research Directions:}
Despite these advancements, several issues remain to be addressed. First, the function class assumption of the ansatz presents a limitation. In this study, the ansatz is confined to the RKHS, which may lead to inaccuracies in the functional gradient direction, as noted in prior work~\cite{dong2022particle}. Second, the iterative procedure assumes a continuity equation, a first-order differential equation with-respect-to time $t$, within the Wasserstein space~\cite{wang2024gad}. However, if the posterior distribution is a ``unbalanced'' multimodal distribution, with some modes having high probability density while others are low, the effectiveness of the proposed InfO-EM algorithm may degrade. In such cases, the underlying PDE should account for particle weights to adequately capture the contribution of each mode~\cite{skretafeynman,wang2025debiased,10795195}. These challenges present important avenues for future research to refine the infinite-horizon-optimal-control-based latent variable inference strategy and improve its robustness and convergence rate.% under more complex distributional with the help of Hamiltonian systems~\cite{7867874}.

\bibliographystyle{IEEEtran}
% Loading bibliography database
\bibliography{ref}

\end{document}